\documentclass[twocolumn,prb,showpacs,multicol,amsmath,amssymb]{revtex4-1}
\usepackage{graphicx}
\usepackage{epstopdf}
\newcommand{\be}{\begin{equation}}
\newcommand{\ee}{\end{equation}}

\newcommand{\bea}{\begin{eqnarray}}
\newcommand{\eea}{\end{eqnarray}}
\newcommand{\bd}{\begin{displaymath}}
\newcommand{\ed}{\end{displaymath}}
\newcommand{\ba}{\begin{array}}
\newcommand{\ea}{\end{array}}
\newcommand{\bi}{\begin{itemize}}
\newcommand{\ei}{\end{itemize}}
\newcommand{\bc}{\begin{center}}
\newcommand{\ec}{\end{center}}
\newcommand{\bfl}{\begin{flushleft}}
\newcommand{\efl}{\end{flushleft}}
\newcommand{\bfr}{\begin{flushright}}
\newcommand{\efr}{\end{flushright}}

\def\6{\partial}

\def\={\!\!\!&=&\!\!\!}
\def\+{\!\!\!&&\!\!\!+~}
\def\-{\!\!\!&&\!\!\!-~}

\begin{document}
 \date{\today}
 \title{RKKY interaction in SDW phase of iron-based superconductors}

\author{Alireza Akbari$^{1,2}$}
\author {Ilya Eremin$^{2}$}
\author {Peter Thalmeier$^{1}$}
 \affiliation{ $^1$Max Planck Institute for the  Chemical Physics of Solids, D-01187 Dresden, Germany\\
 $^{2}$ Institut f\"ur Theoretische Physik III, Ruhr-Universit\"{a}t Bochum, 44801 Bochum, Germany
 }
 \begin{abstract}
Using the multiband model we analyze the Ruderman-Kittel-Kasuya-Yosida (RKKY) interaction between the magnetic impurities in layered ferropnictide superconductors. In the normal state the interaction is spin isotropic and is dominated by the nesting features of the electron and hole bands separated by the antiferromagnetic momentum, ${\bf Q}_{AF}$.
In the AF state the RKKY interaction maps into an effective anisotropic XXZ-type Heisenberg exchange model. The anisotropy originates from the breaking of the spin-rotational symmetry induced by the AF order and its strength depends on the size of the AF gap and the structure of the folded Fermi surface. We discuss our results in connection to the recent experiments. \end{abstract}

 \maketitle
 \section{Introduction}

The oscillatory Ruderman-Kittel-Kasuya-Yosida (RKKY) interaction\cite{Ruderman54,Kasuya56,Yosida57}
of the localized magnetic moments in a metal has always played
an important role in revealing the nature of the magnetic interaction in metals with partially unfilled $d$- and $f$-electron shells.
This indirect interaction is the result of the spin polarization of conduction
electrons produced by the exchange interaction of the localized moments with
conduction electrons where the distance between two
localized moments controls the strength of their effective exchange.
Originally formulated for the three-dimensional spherical Fermi surface, RKKY interaction has been also analyzed for the two-dimensional (2D)\cite{Fisher75,Monod87}
as well as one-dimensional (1D)\cite{Yafet87} electron gas. A particular interesting situation arises in
a highly anisotropic Fermi surface with nesting. In particular, it was shown that the RKKY interaction in such a case consists of several terms originating from flat regions in the
fermionic spectrum (van Hove regions) and those which describe the interference
between contributions from their vicinities. \cite{Aristov97a}
Note that
the latter terms have an overall prefactor
$\cos({\bf Q \cdot r})$ with {\bf Q} = $(\pi,\pi)$. In the
nearly nested situation this last term is present down to the
interatomic distances and favors the commensurate
antiferromagnetic ordering of the localized moments.

It is important to bear in mind that in order to evaluate the RKKY interaction
at the distances $r$ the details of the
fermionic dispersion on a scale of $1/r$ in the $k$-space is required. As a result the
fine details of the Fermi surface are only important at
largest distances while even the approximate
knowledge of the electronic spectrum over the whole Brillouin zone is often
enough for a good description of the RKKY interaction at the interatomic distances in
$r$ space. In this regard it is quite instructive to analyze the RKKY interaction in recently discovered
Fe-based superconductors\cite{Kamihara08}.
Band structure calculations\cite{LDA} and experimental probes such as angle-resolved photoemission (ARPES)\cite{ARPES} and quantum oscillation\cite{coldea,sebastian} experiments show that to a good approximation
the Fermi surface topology of iron-based superconductors consists of the small sized circular hole
pockets centered around the $\Gamma-$point $(0,0)$, and  elliptic electron
pockets centered around the $(\pm\pi,0)$, and $(0,\pm\pi)$-points of the unfolded Brillouin zone (BZ).
The pockets are nearly of the same size which results in the
nesting properties of the electron and hole bands at wave vectors,
${\bf Q}_{i}$ ( ${\bf Q}_{1}=(\pi,0)$, and  ${\bf Q}_{2}=(0,\pi)$),
 {\it i.e.}
  $\varepsilon_{\bf k}^{e}\simeq -\varepsilon_{\bf k+Q_{i}}^{h}$.
Given the electronic structure of ferropnictides, it is natural to assume that
magnetic order emerges, at least partly,  due to near-nesting between
the dispersions of holes and electrons~\cite{Tesanovic,Chubukov2008,d_h_lee,Korshunov2008,timm,honerkamp,eremin}.

\begin{figure}[h]
 \centering
\includegraphics[angle=0,width=.7\linewidth]{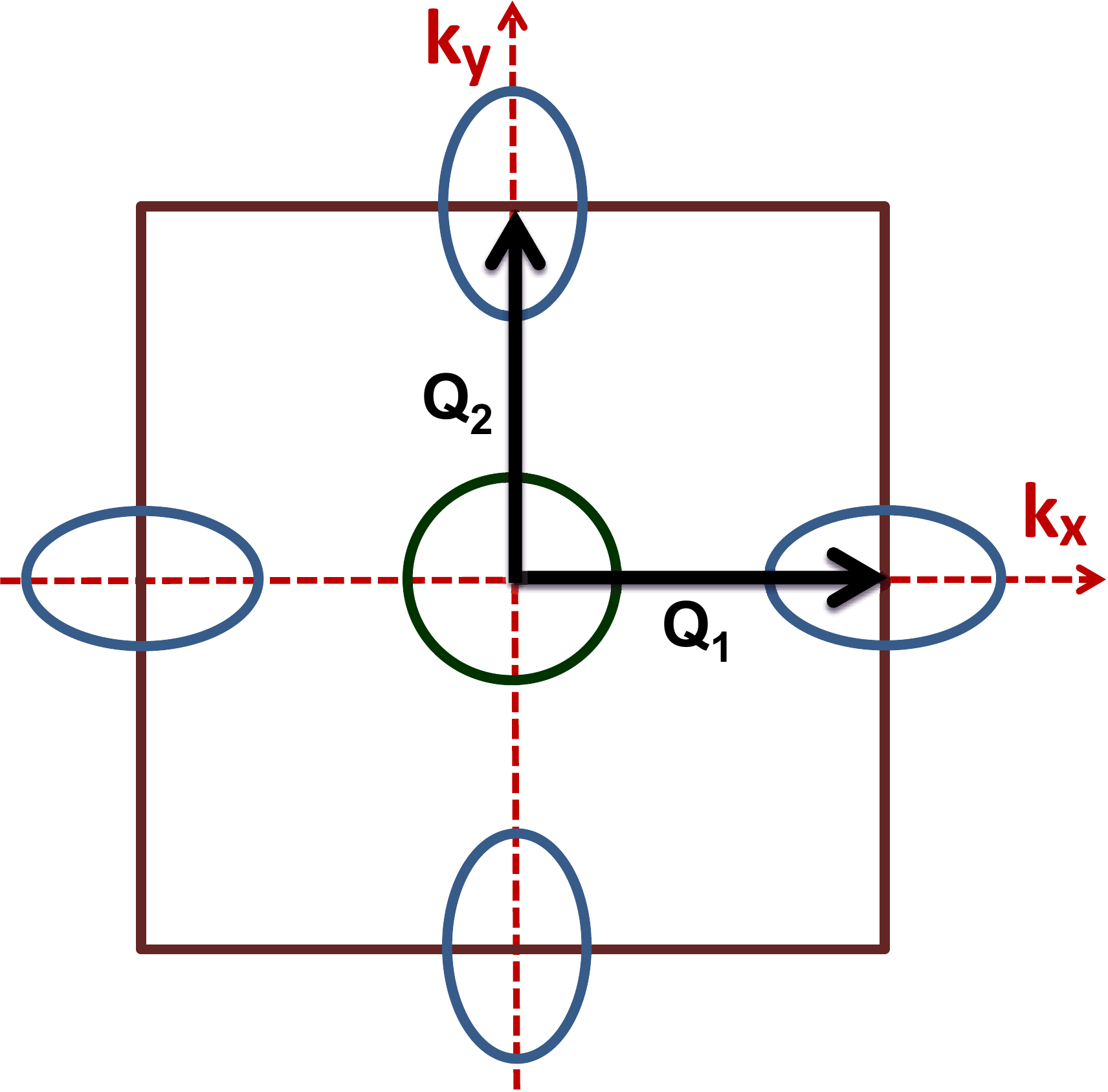}
\caption{(color online) Schematic Fermi surface topology of iron-based
superconductors in the unfolded Brillouine zone (BZ) with 1 Fe per unit cell containing circular hole pocket centered around the $\Gamma-$point
and two electron pockets centered around the $(\pi,0)$ and $(0,\pi)$ points, respectively.}
\label{fermi}
\end{figure}

%
\begin{figure*}
\centerline{
a)\includegraphics[width=0.2\linewidth,angle=0]{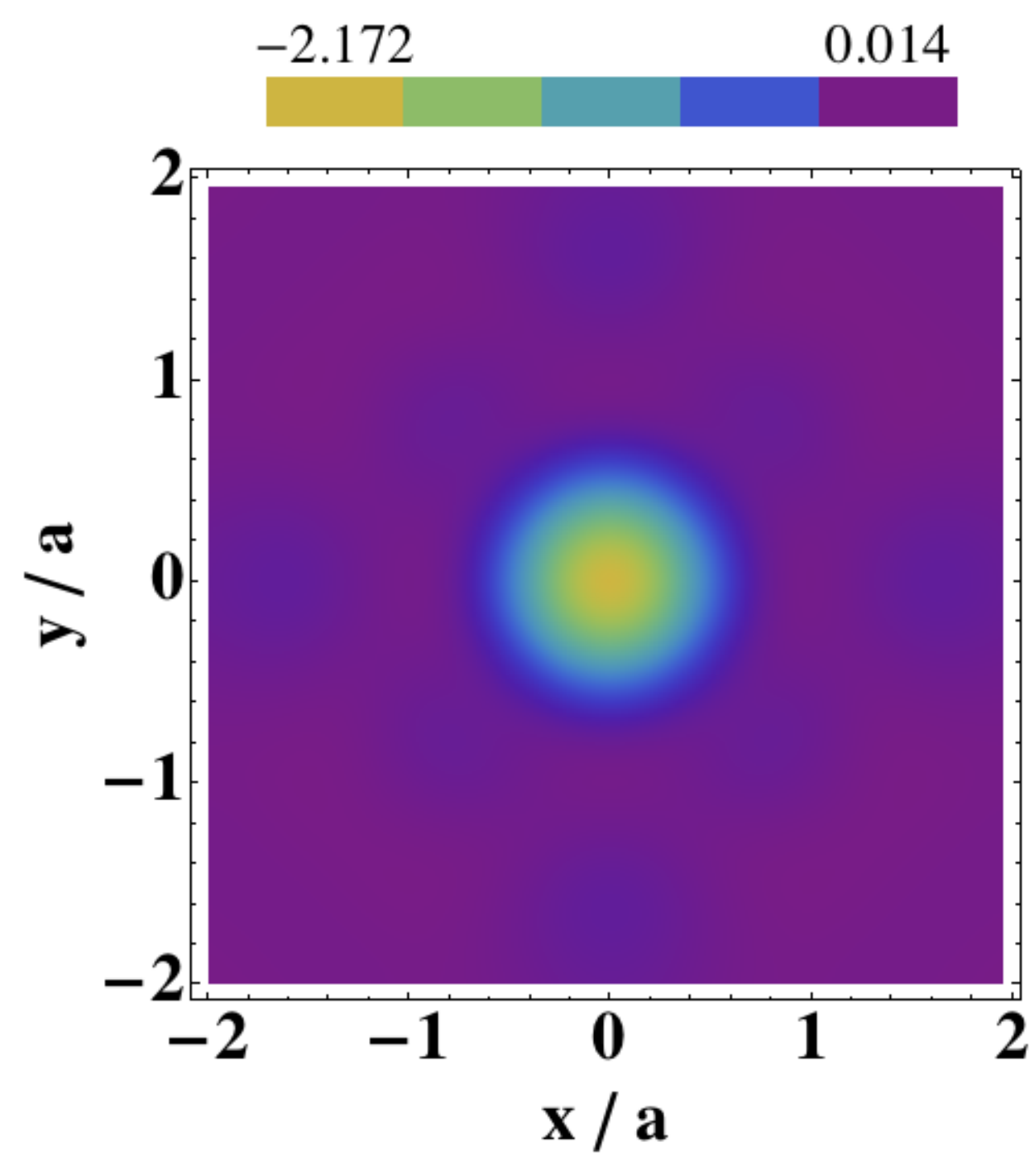}
b)\includegraphics[width=0.2\linewidth,angle=0]{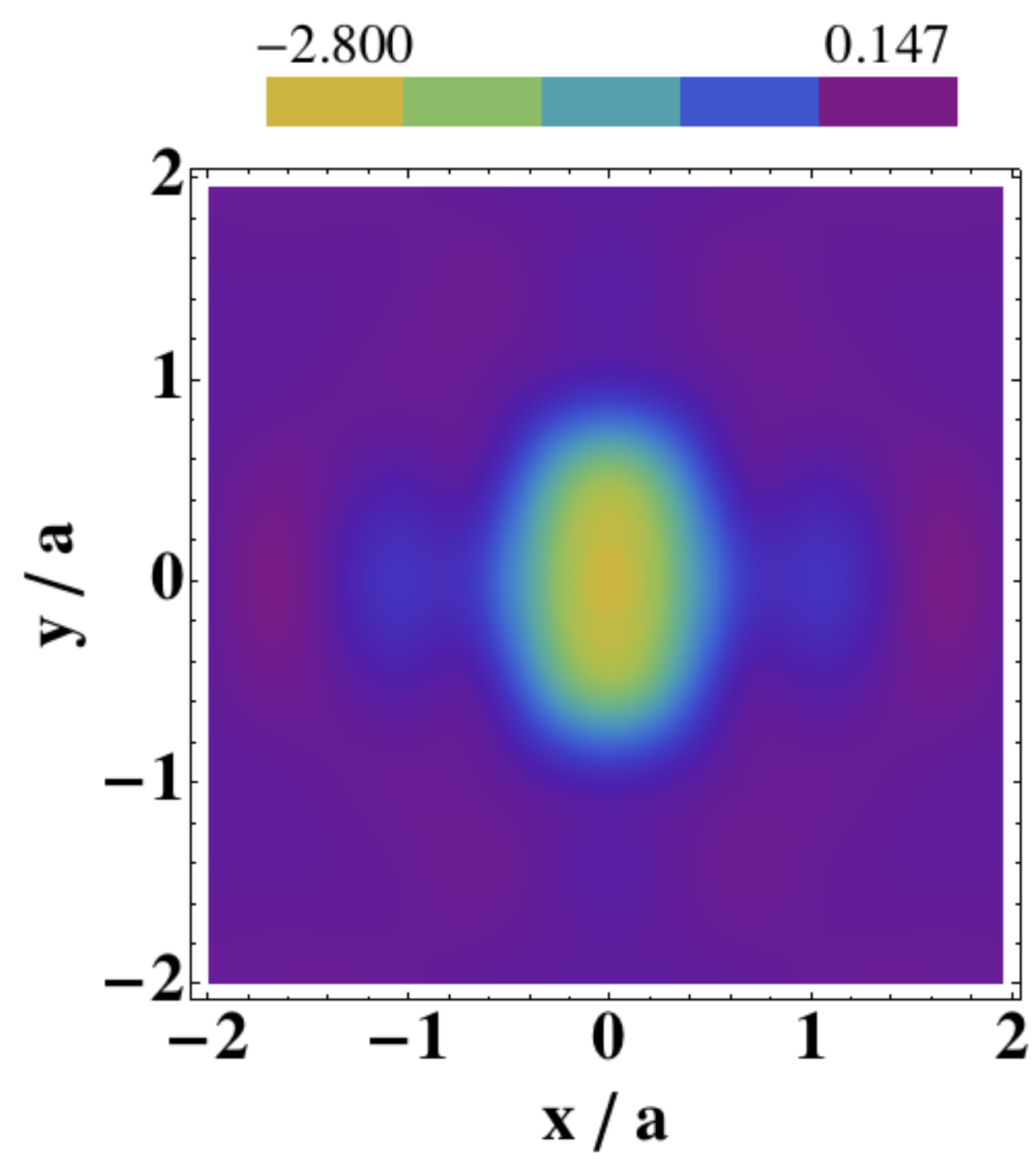}
c)\includegraphics[width=0.2\linewidth,angle=0]{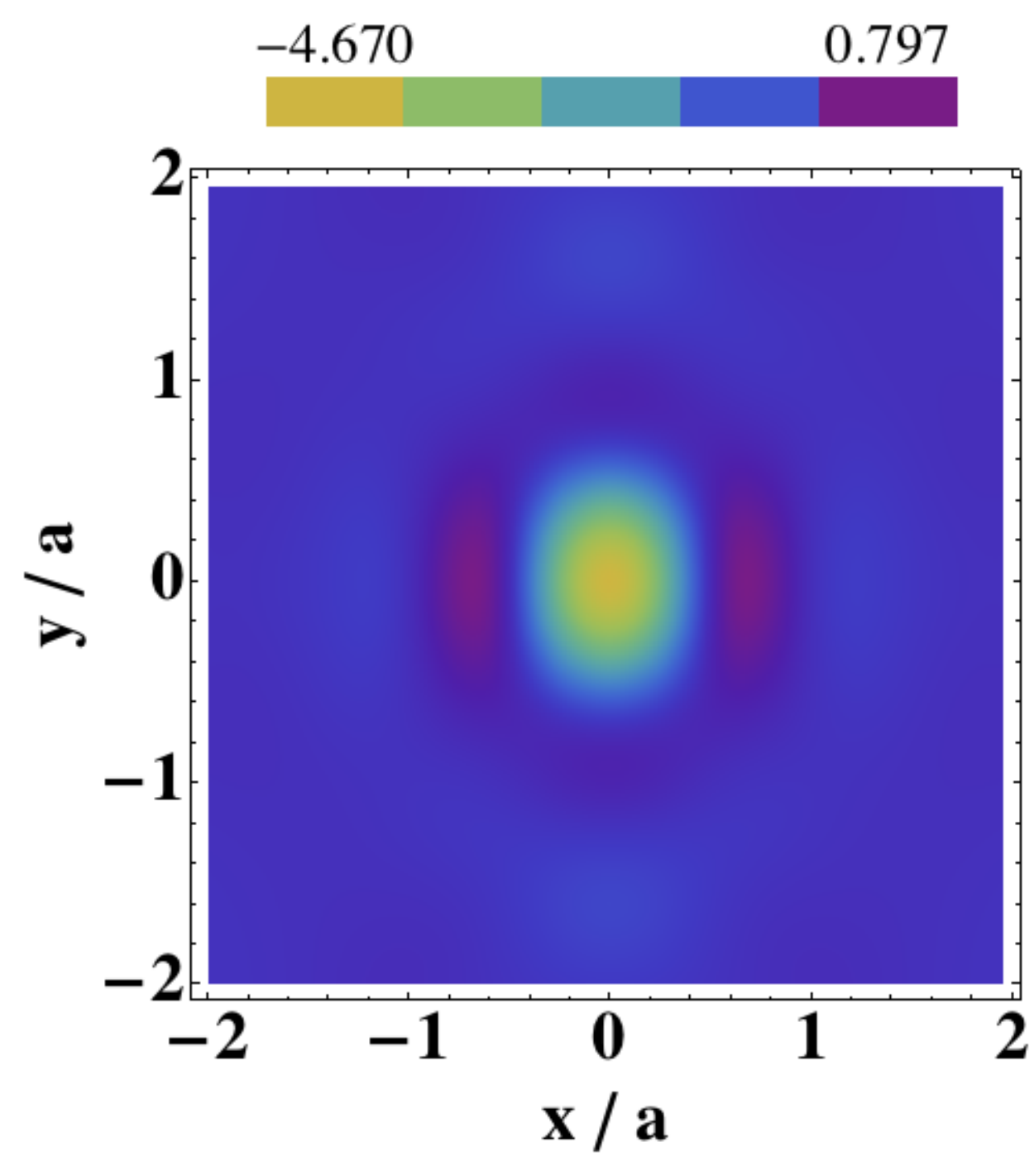}}

\caption{ (color online)
The contour mesh of the RKKY interaction for the normal (a) and the SDW state for $J_x$  (b), and $J_z$ interaction (c).
In the SDW state we employ $W=40meV$, $\epsilon=0.5$, and $5\%$ of the electron doping.}
\label{fig2}
\end{figure*}

Here we analyze the novel aspects of the RKKY
interaction in iron-based superconductors which arise due to the peculiar Fermi surface topology in these systems.
The origin of the local moments in ferropnictides can be either 4$f$-electrons in ReFeAsO series (Re- is a rare-earth element)\cite{maeter,pourovskii} or possibly partially localized $d$-electrons which arise due to proximity to a Mott insulator\cite{si} which coexist with itinerant ones.
Our primary interest is to investigate the evolution of the oscillatory behavior of the RKKY interaction in the presence of nesting in the normal and in the antiferromagnetic states of iron-based superconductors.
Analyzing the RKKY interaction in the antiferromagnetic state we find spin space anisotropy of the interaction, a feature that has not been reported so far.
We will also study in detail the influence of the model parameters like
ellipticity of the electron pockets, and SDW gap size on the spatial variation
of the RKKY interaction.

The paper is organized as follows In Sec. II we evaluate the RKKY interaction for a three band model and
present its analytical form for SDW and normal state regimes.
Using these results we evaluate the RKKY interaction numerically in Sec. III and
discuss its relevance for the experiments. We finally present a summary and conclusion in Sec. IV.

\section{Multi band RKKY interaction in the normal and spin density wave states}

In this investigation we employ a minimal model of
interacting 3d electrons and local moments in ferropnictides with a circular 3$d$ hole Fermi surface (FS)
centered around  $\Gamma$-point ($a$-band)
and two elliptical electron FS pockets centered around $(\pm \pi, 0)$ and $(0, \pm \pi)$ points in the unfolded BZ ($b$-bands) (See Fig.\ref{fermi}).


%
\begin{figure*}
\centerline{
a)\includegraphics[width=0.2\linewidth,angle=0]{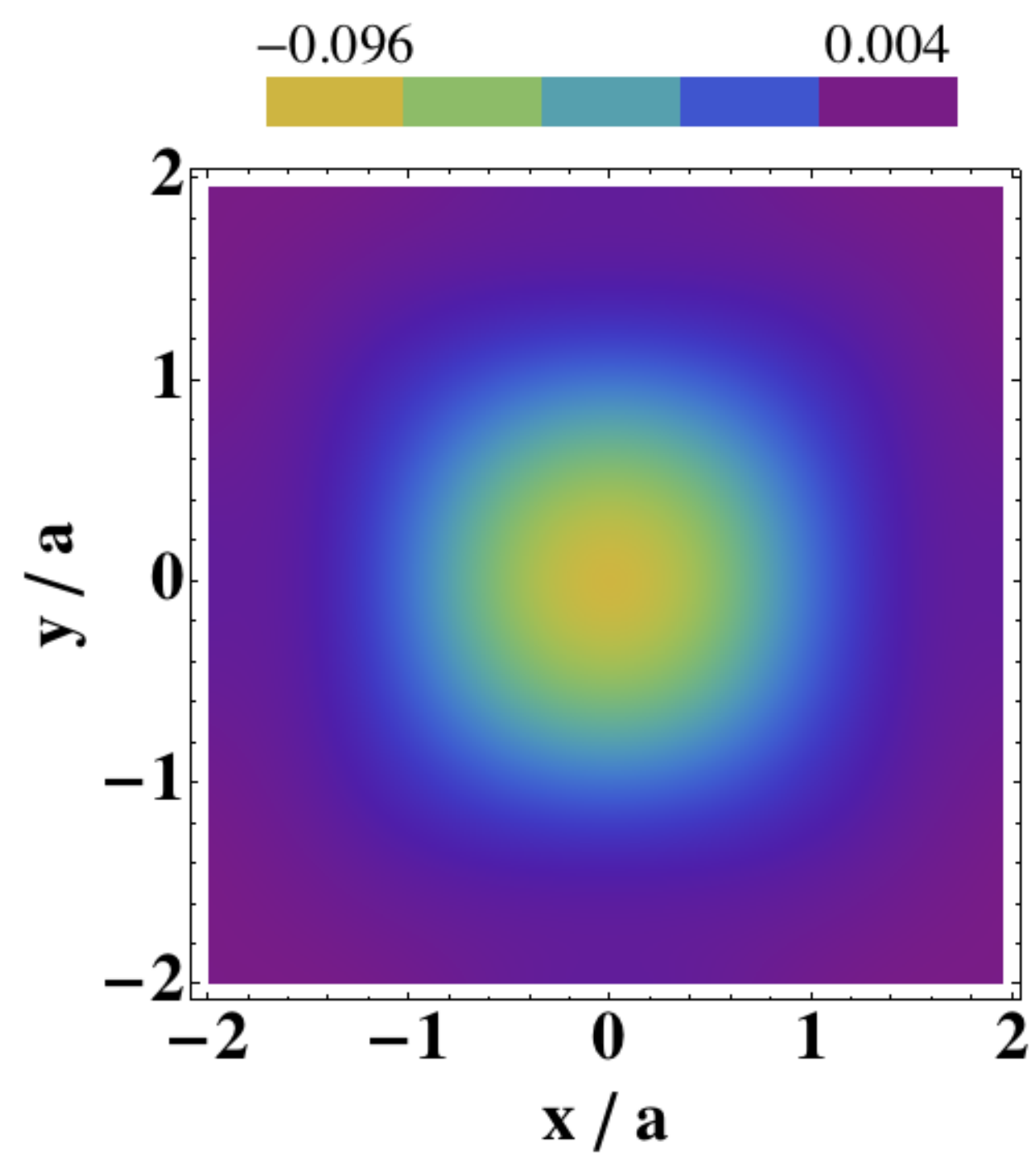}
b)\includegraphics[width=0.2\linewidth,angle=0]{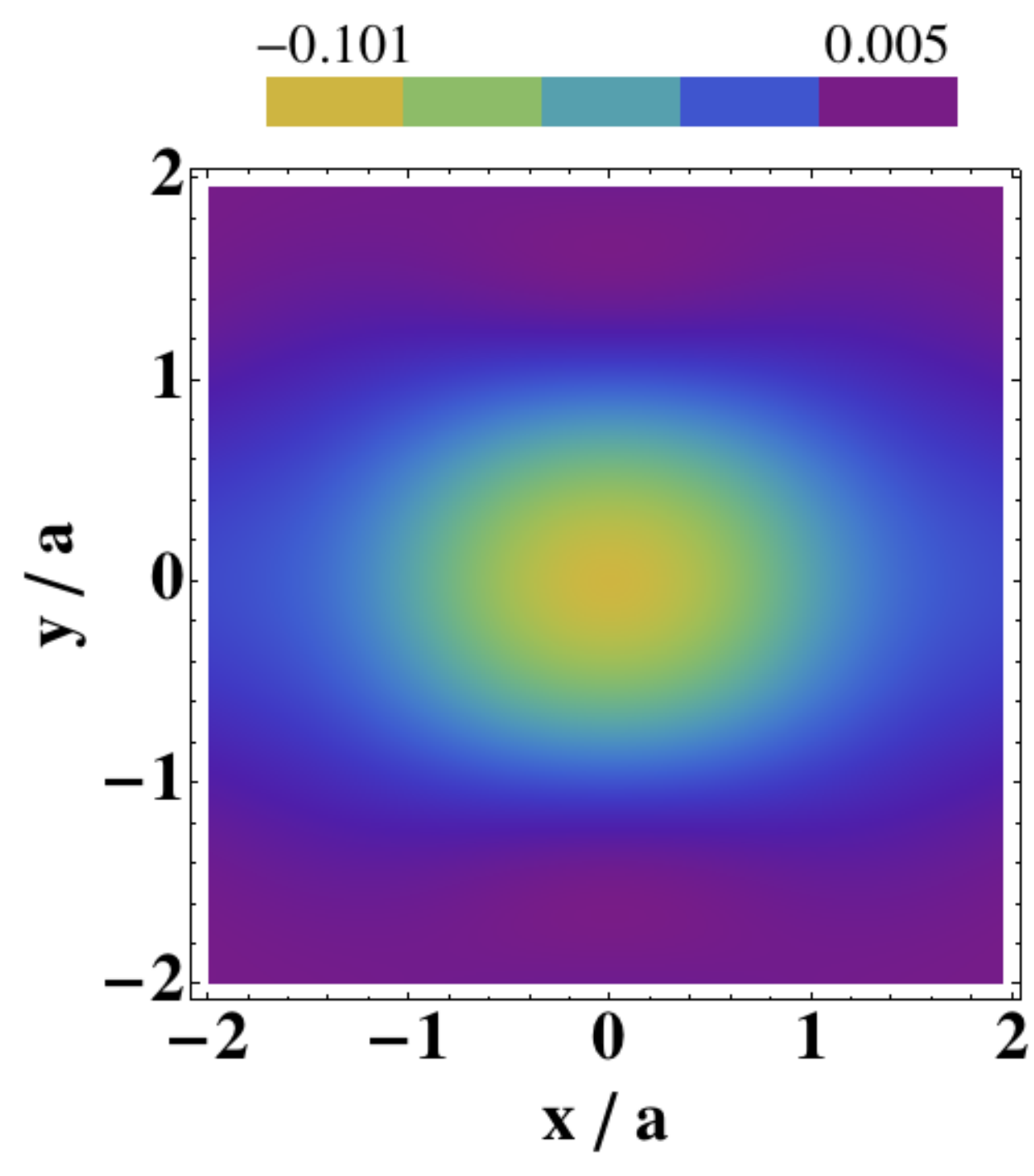}
c)\includegraphics[width=0.2\linewidth,angle=0]{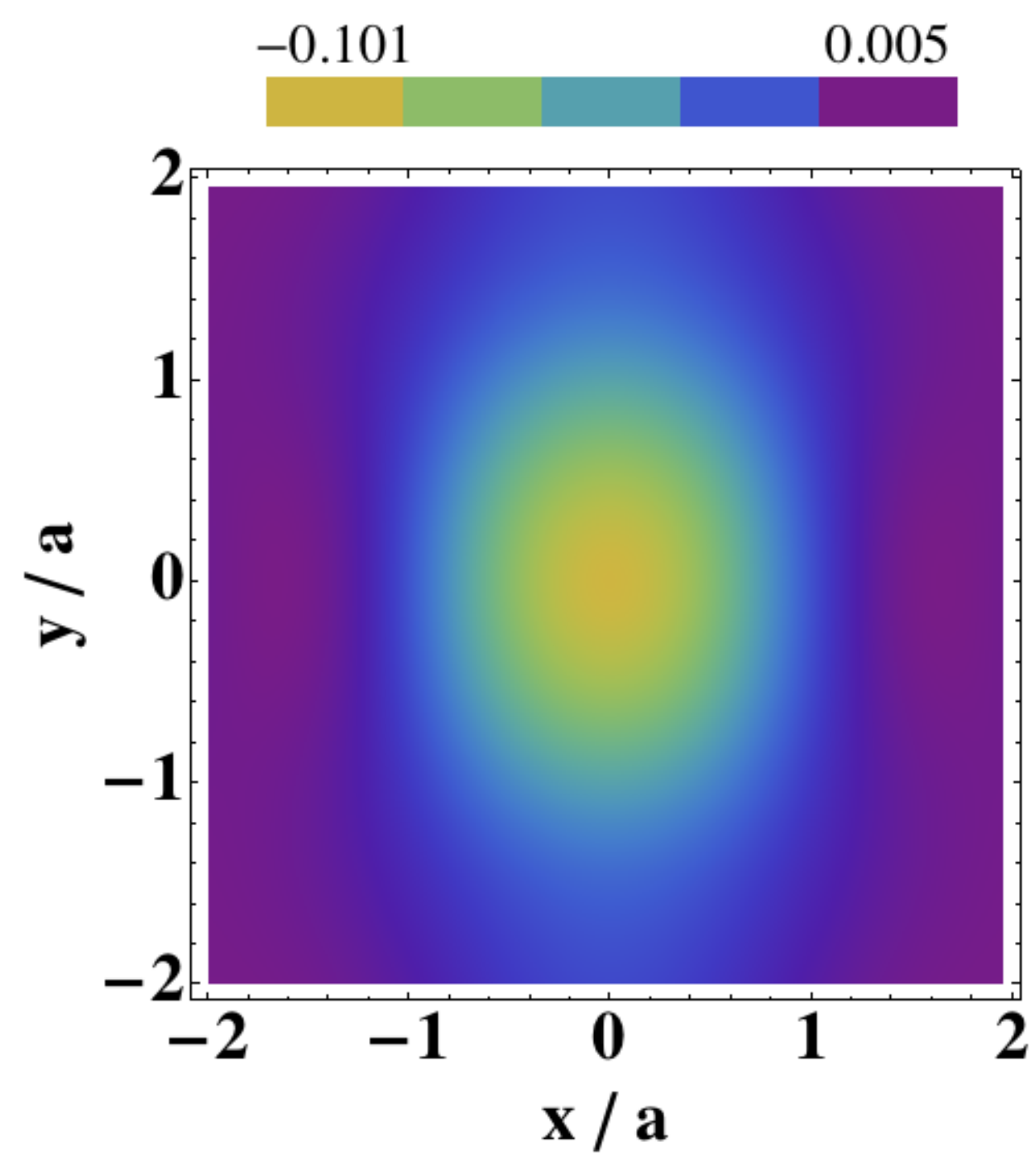}}

\centerline{
d)\includegraphics[width=0.2\linewidth,angle=0]{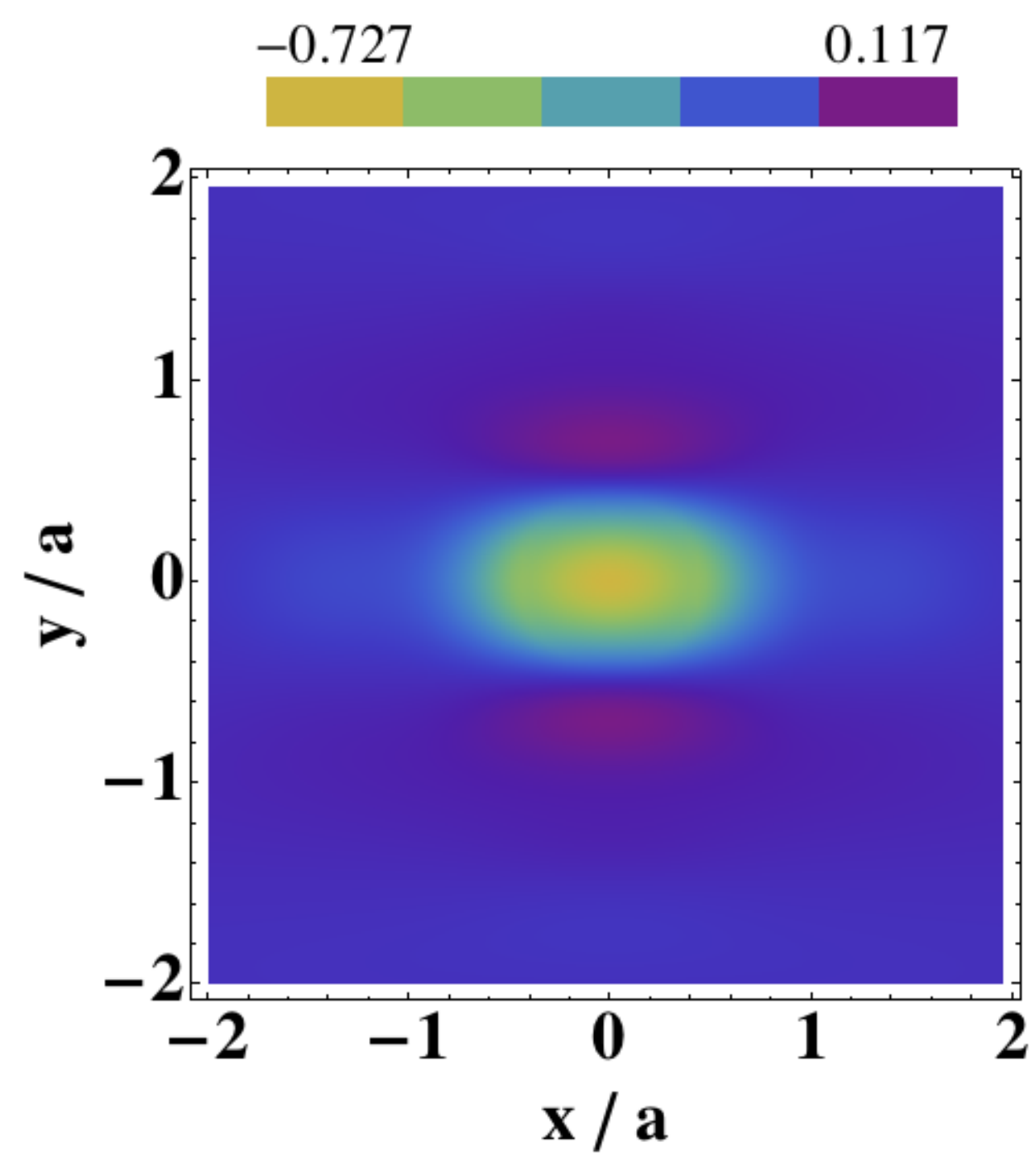}
e)\includegraphics[width=0.2\linewidth,angle=0]{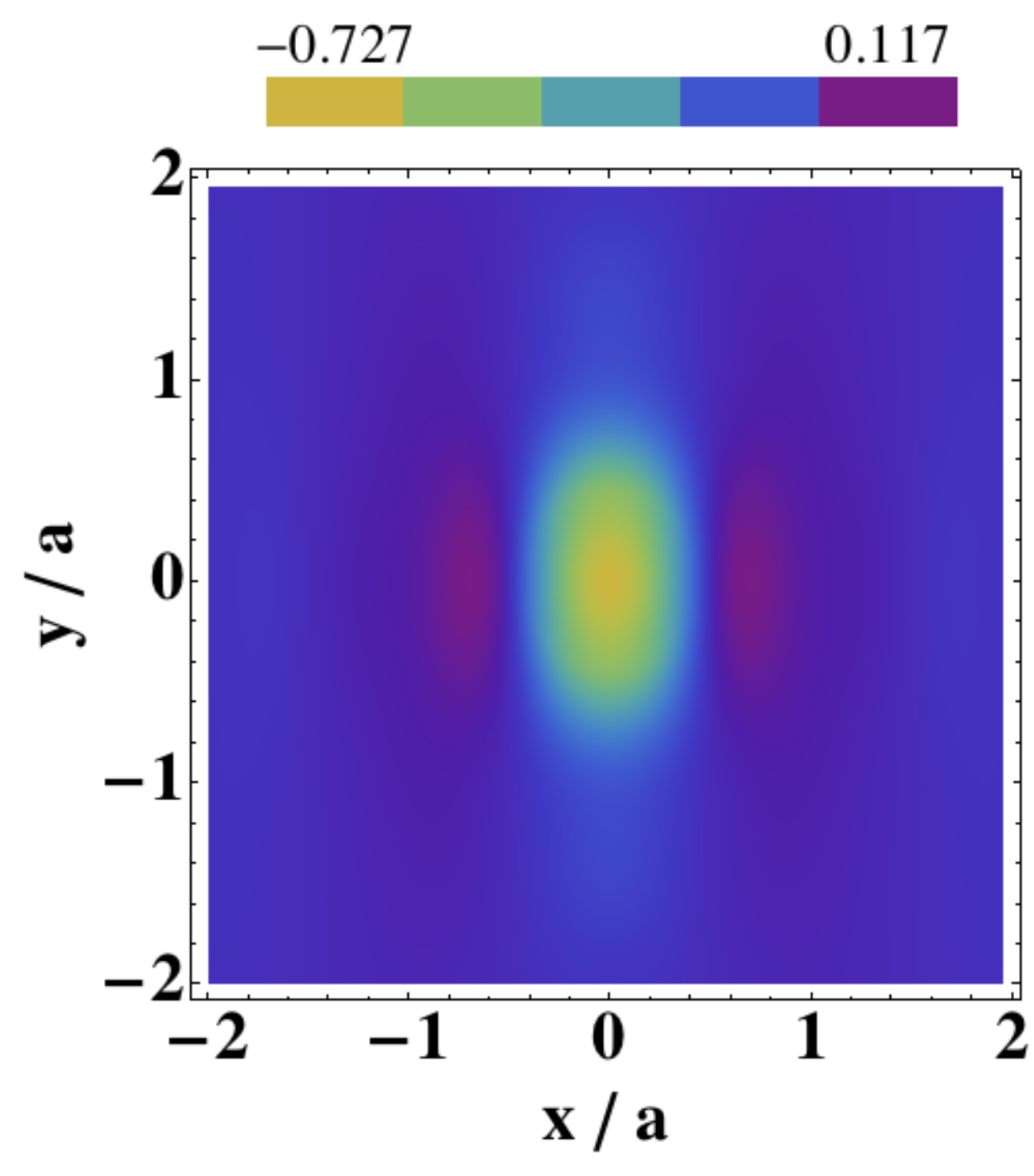}
f)\includegraphics[width=0.2\linewidth,angle=0]{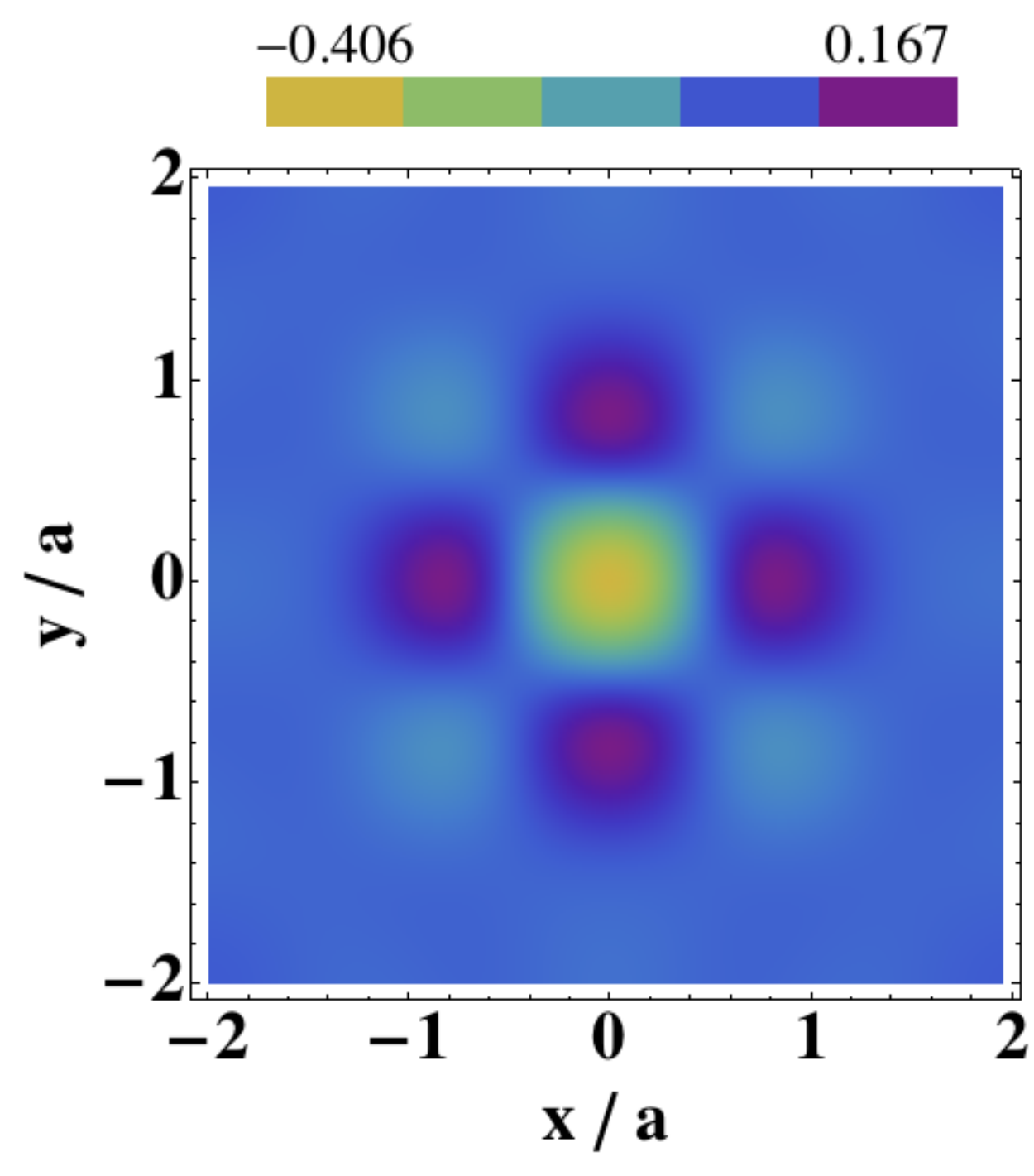}}
\caption{(color online)
The intra-band (upper panel, (a) ${\cal J}_n^{11}$, (b) ${\cal J}_n^{22}$, (c) ${\cal J}_n^{33}$) and inter-band  (lower panel (d) ${\cal J}_n^{12}$, (e) ${\cal J}_n^{13}$, and (f) ${\cal J}_n^{23}$) contributions to the RKKY interaction in the normal state for $\epsilon=0.5$ and $5\%$  of the electron doping.}
\label{fig3}
\end{figure*}
\begin{figure*}
\centerline{
a)\includegraphics[width=0.2\linewidth,angle=0]{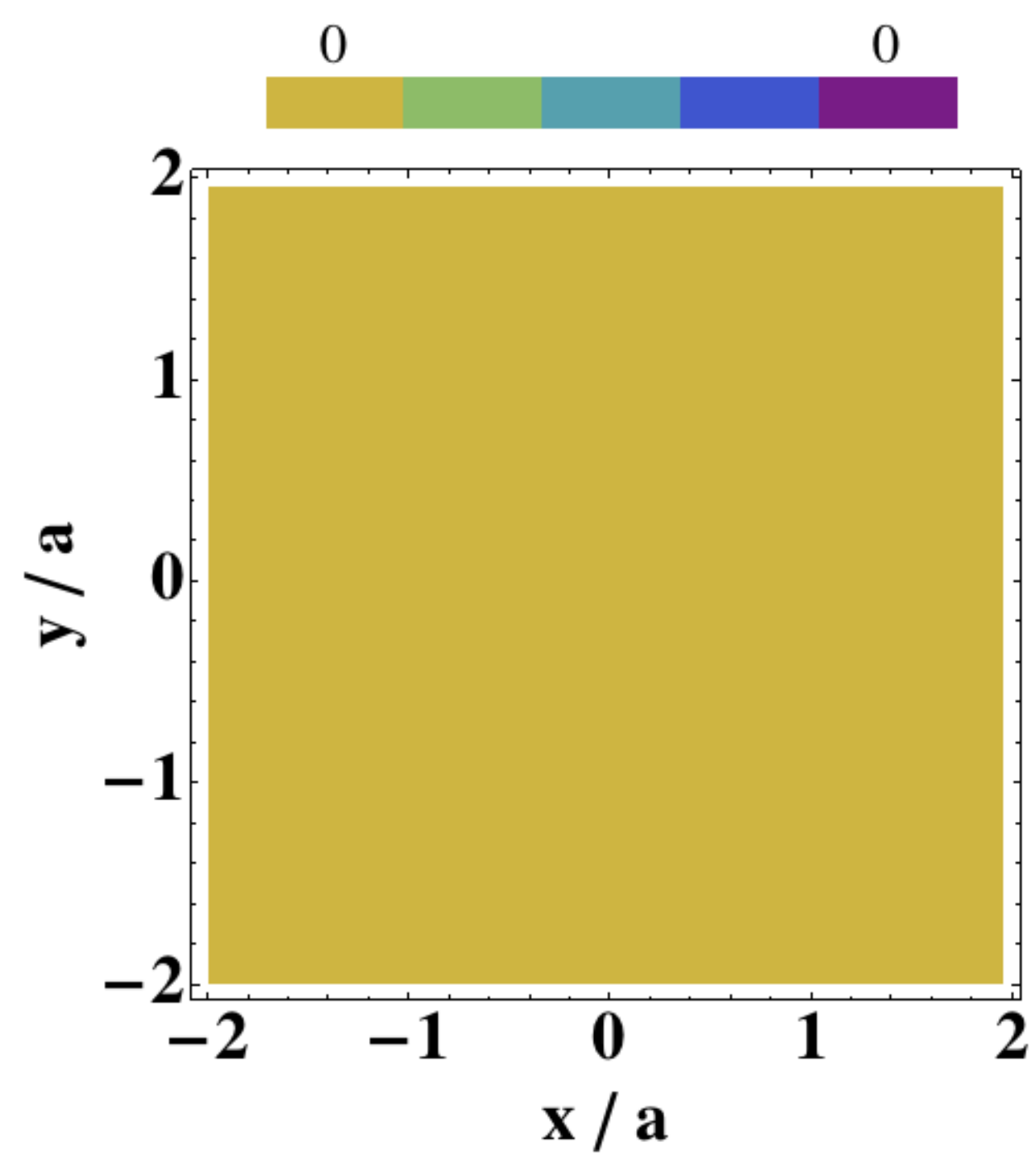}
b)\includegraphics[width=0.2\linewidth,angle=0]{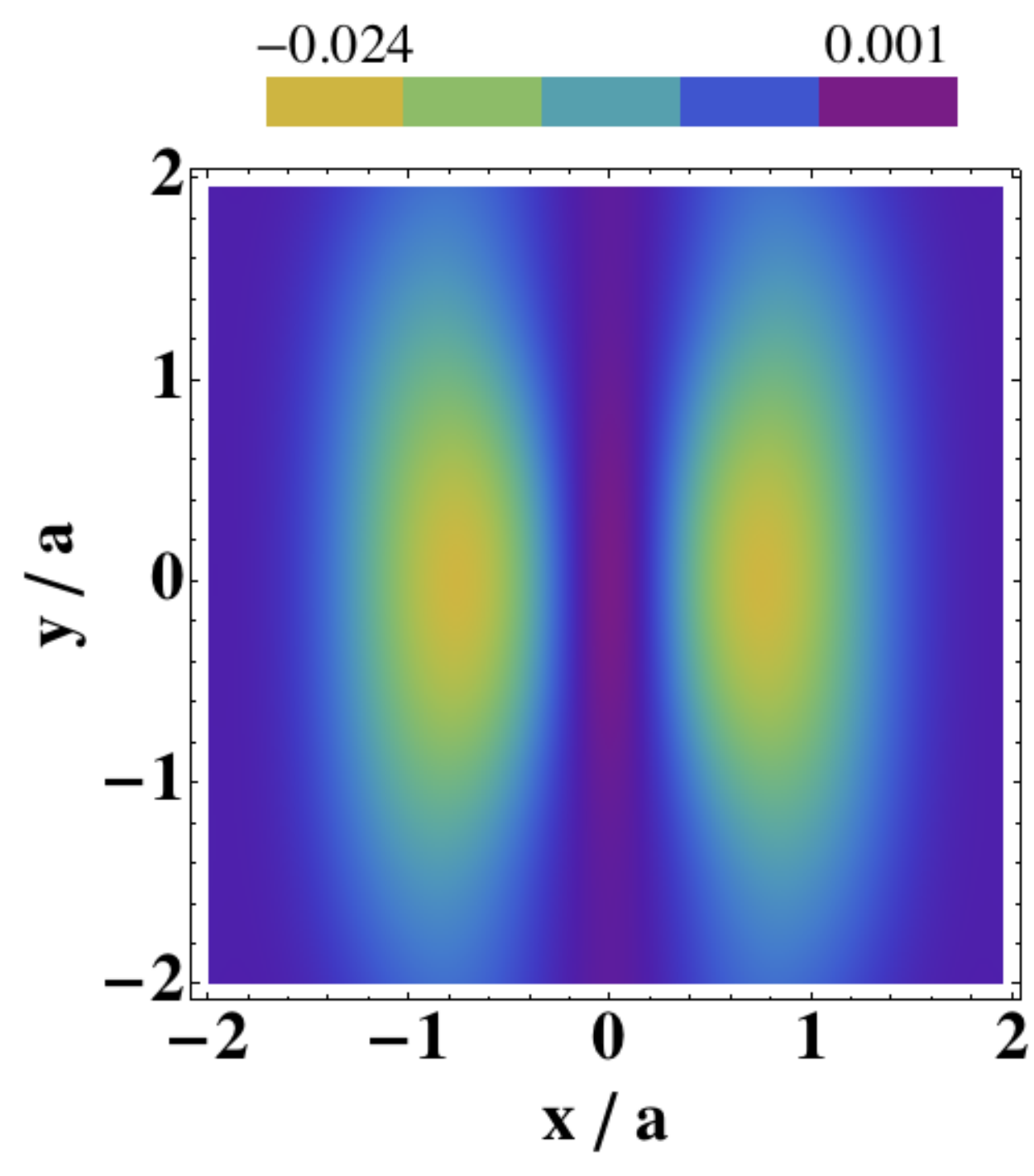}
c)\includegraphics[width=0.2\linewidth,angle=0]{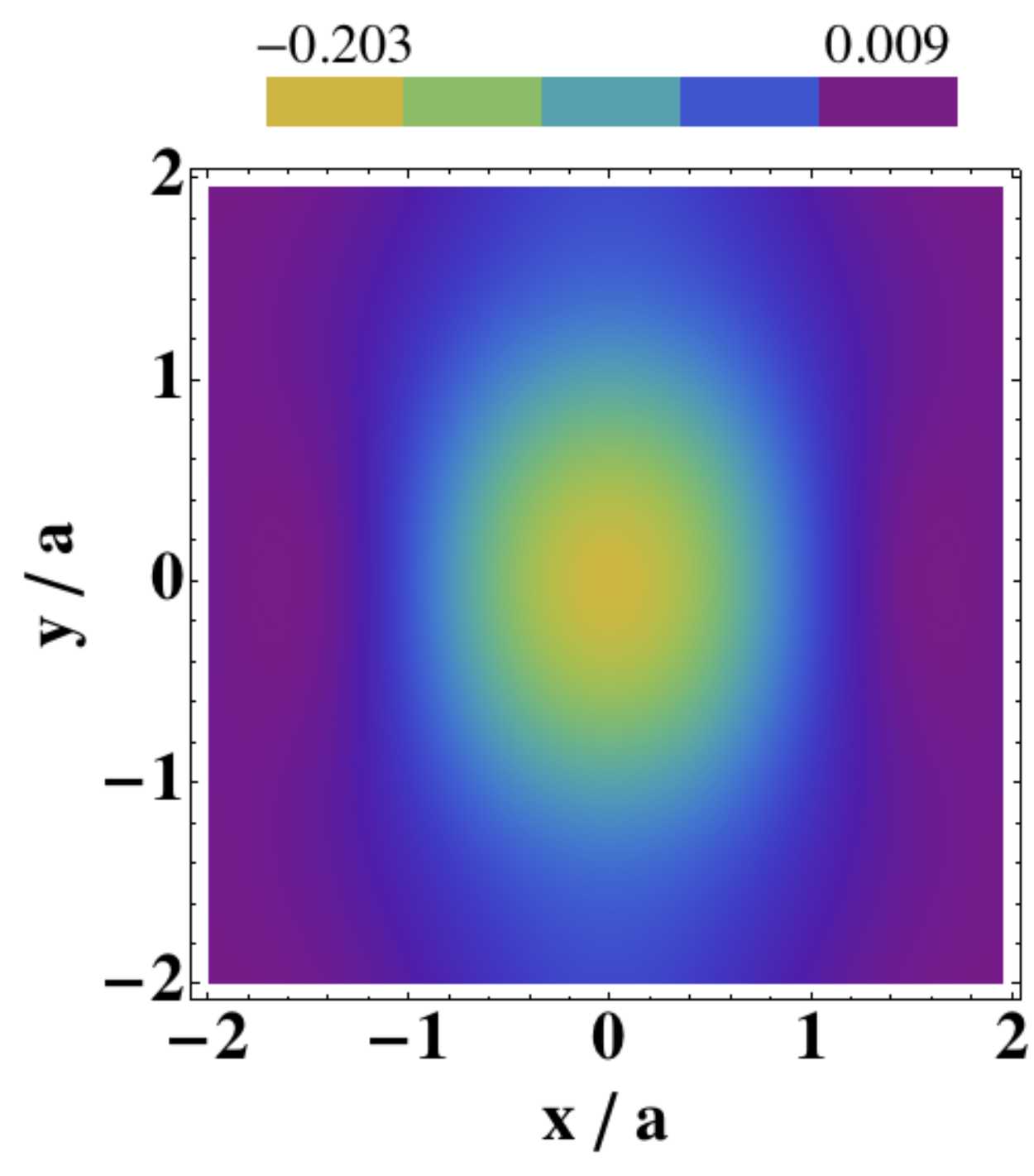}}

\centerline{
d)\includegraphics[width=0.2\linewidth,angle=0]{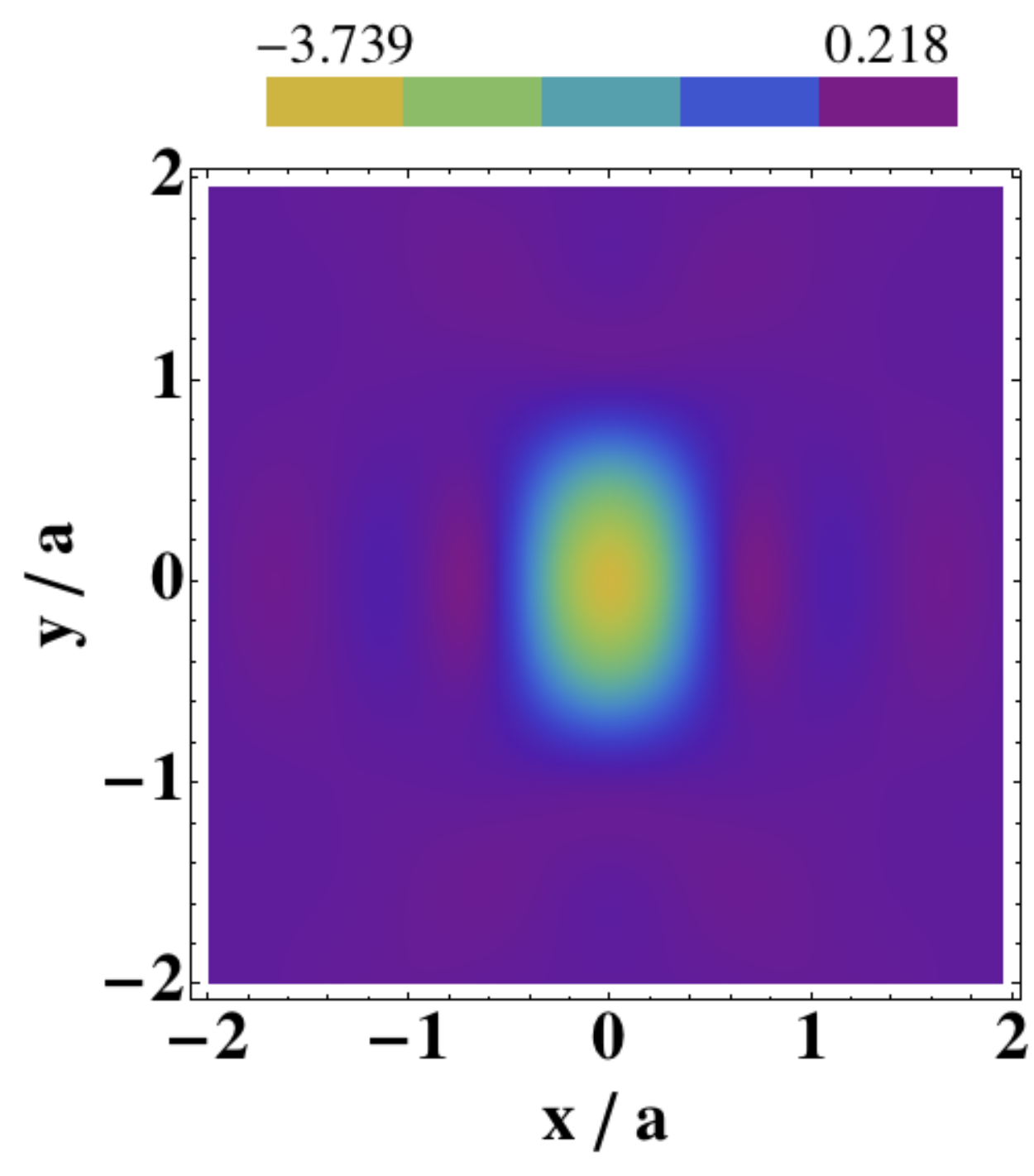}
e)\includegraphics[width=0.2\linewidth,angle=0]{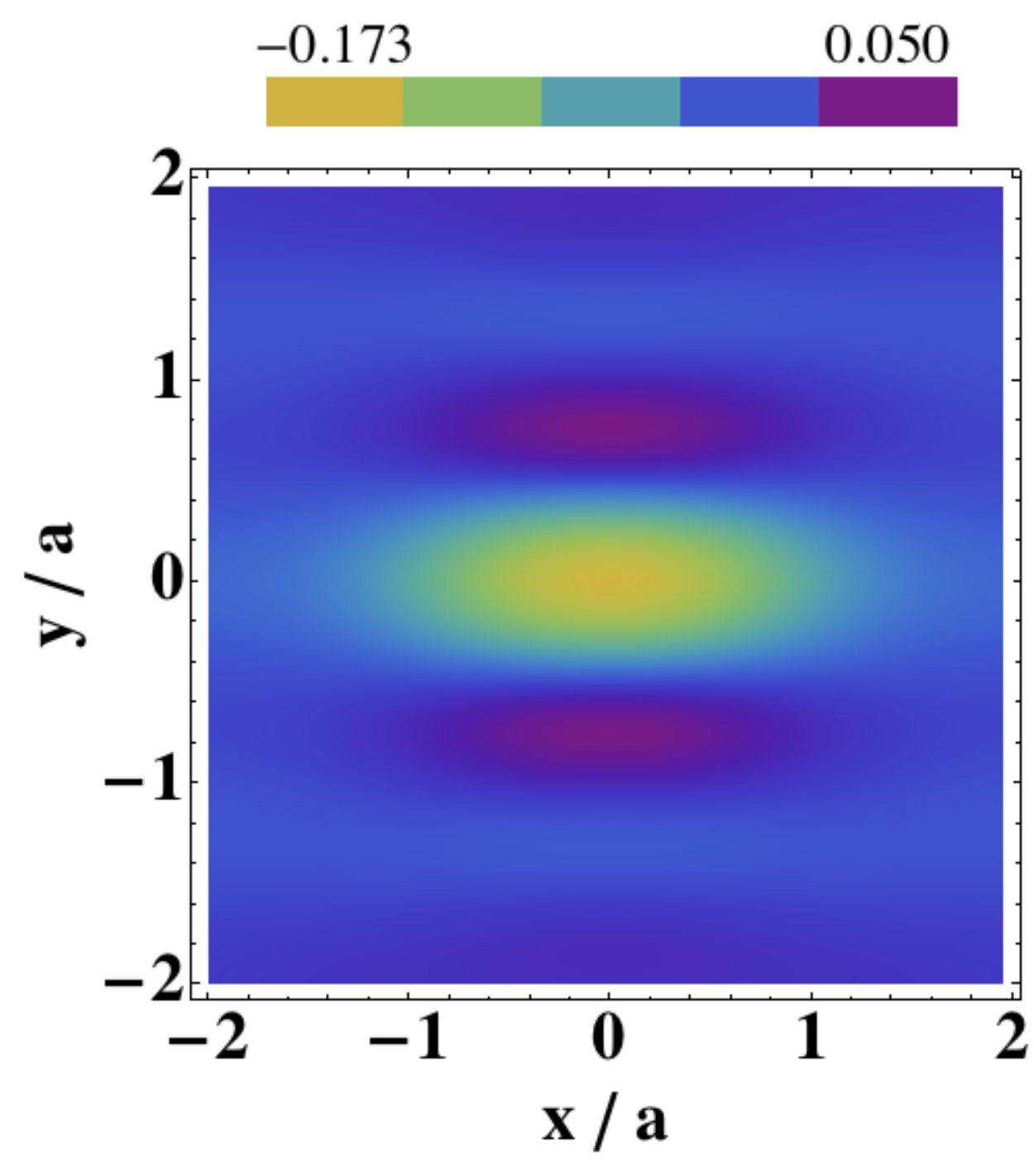}
f)\includegraphics[width=0.2\linewidth,angle=0]{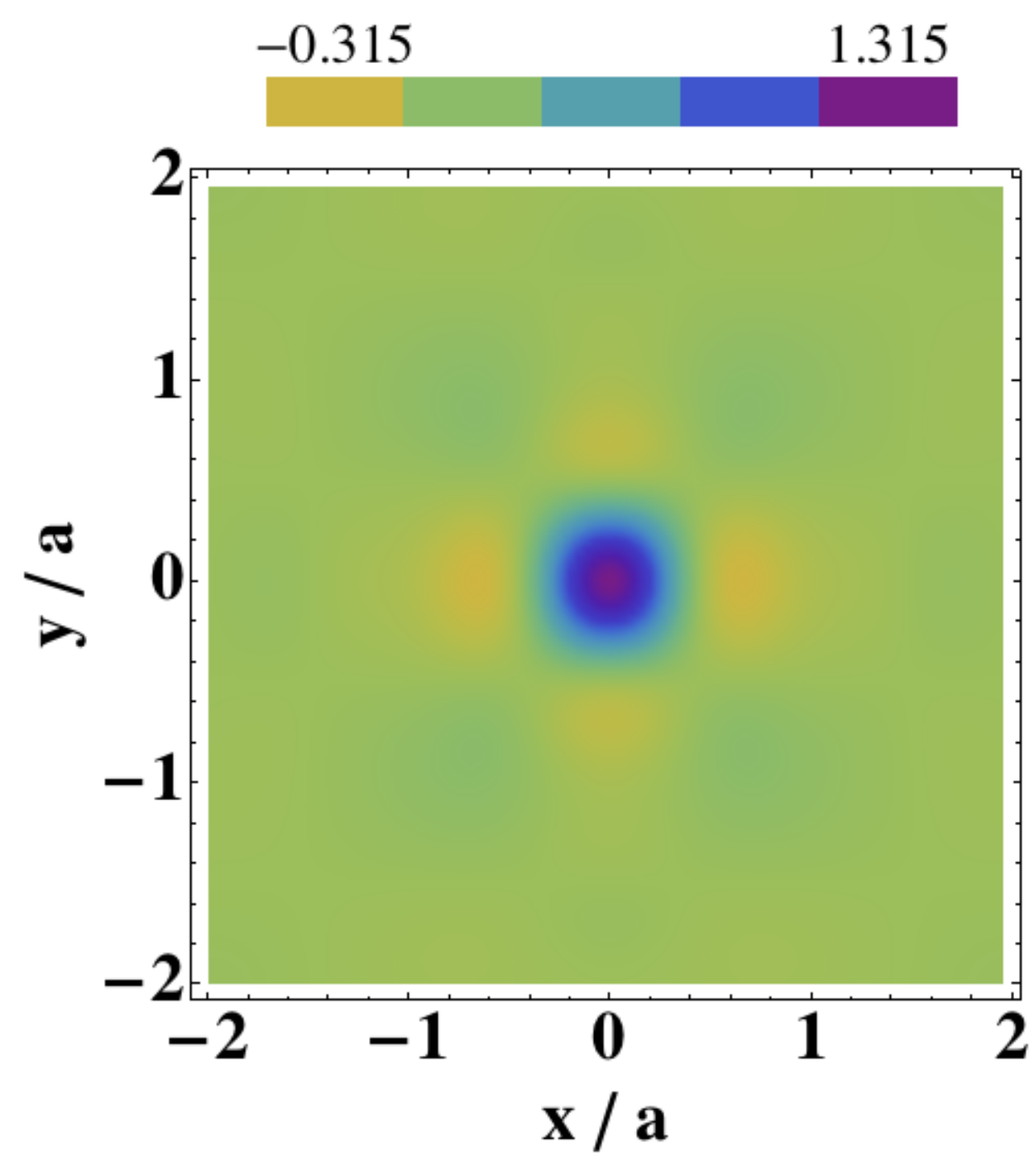}}
\caption{(color online)
The intra-band (upper panel (a) ${\cal J}_x^{11}$, (b) ${\cal J}_x^{22}$, (c) ${\cal J}_x^{33}$) and interband (lower panel (d) ${\cal J}_x^{12}$, (e) ${\cal J}_x^{13}$, and (f) ${\cal J}_x^{23}$) contributions to the RKKY interaction, $J^x$ in SDW state
 $x$-direction for $W=40meV$, $\epsilon=0.5$, and $5\%$ of the electron doping.}
\label{fig4}
\end{figure*}

\begin{figure}
\centerline{
\includegraphics[width=0.8\linewidth,angle=0]{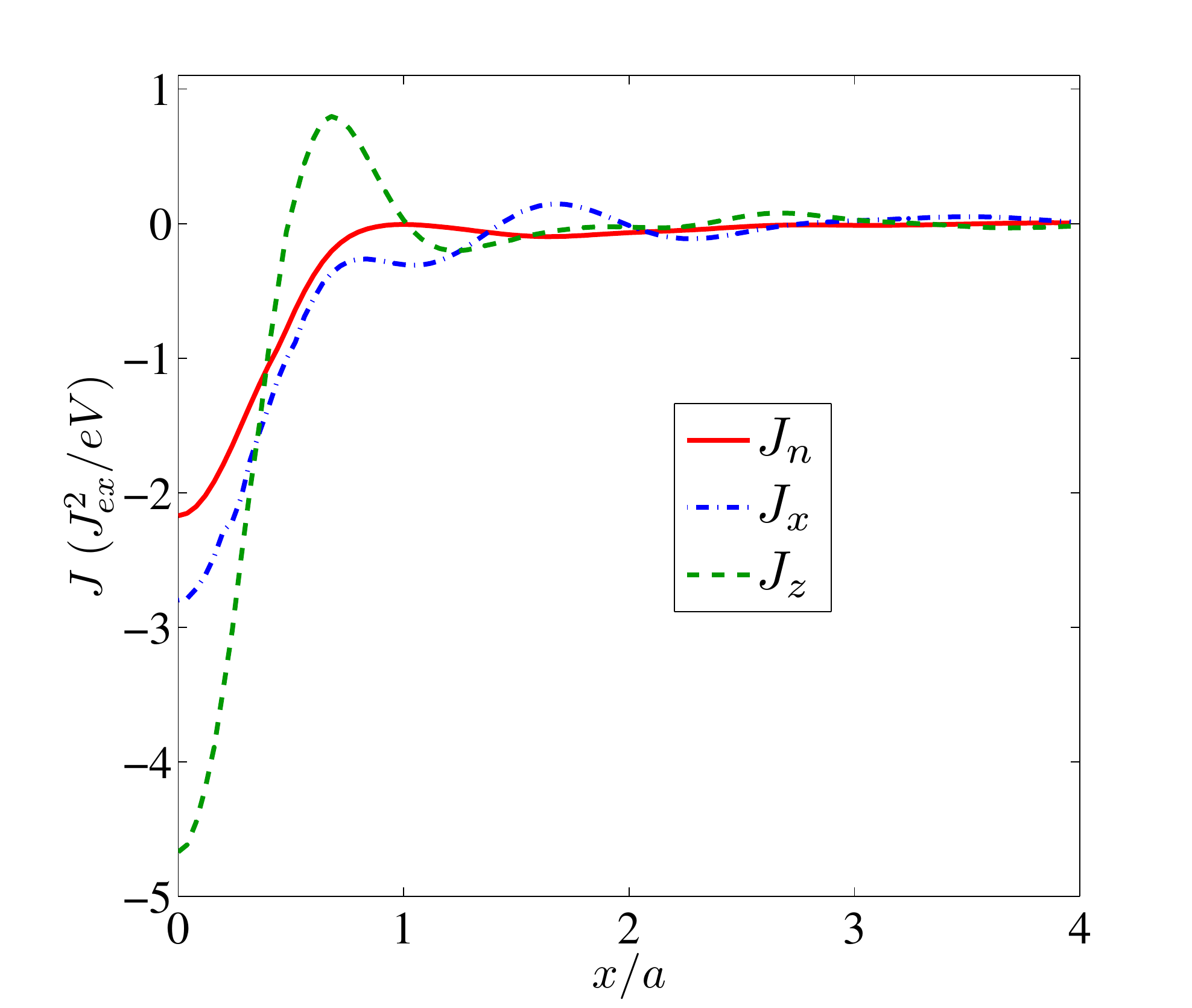}}
\caption{  (color online)
Oscillatory behavior of the RKKY interaction along $x$ direction in the normal ($J_n$)
and SDW ($J_x$ and $J_z$) states. }
\label{fig5}
\end{figure}
\begin{figure}
\centerline{\includegraphics[width=0.8\linewidth,angle=0]{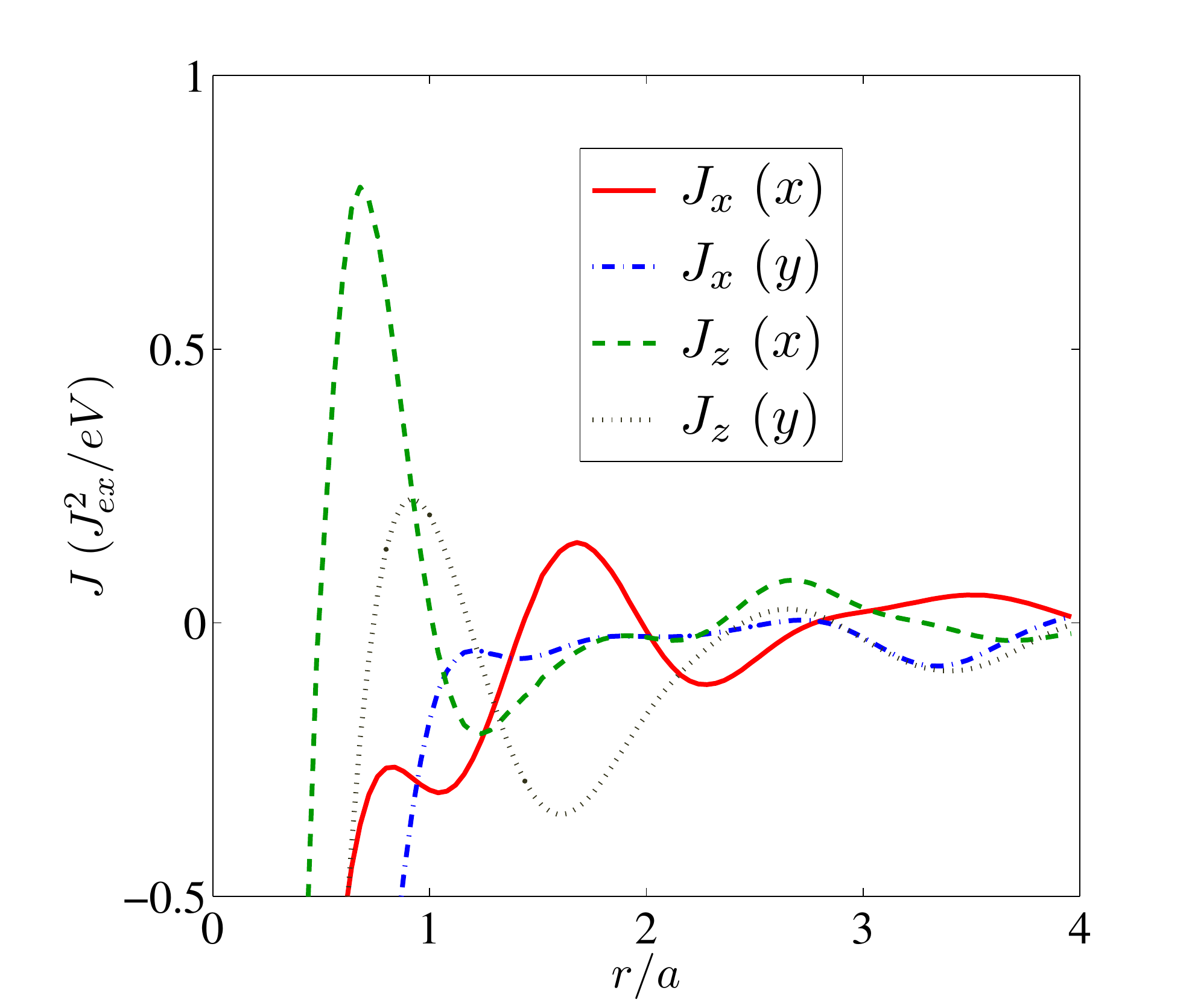}}
\caption{ (color online)
$J_x$ and $J_z$ components of the XXZ RKKY interaction in the SDW state along two crystallographic directions. As above we employ
$W=40meV$, $\epsilon=0.5$, and $5\%$ e-doping.
 }
\label{fig6}
\end{figure}
\begin{figure*}\centerline{
a)\includegraphics[width=0.4\linewidth,angle=0]{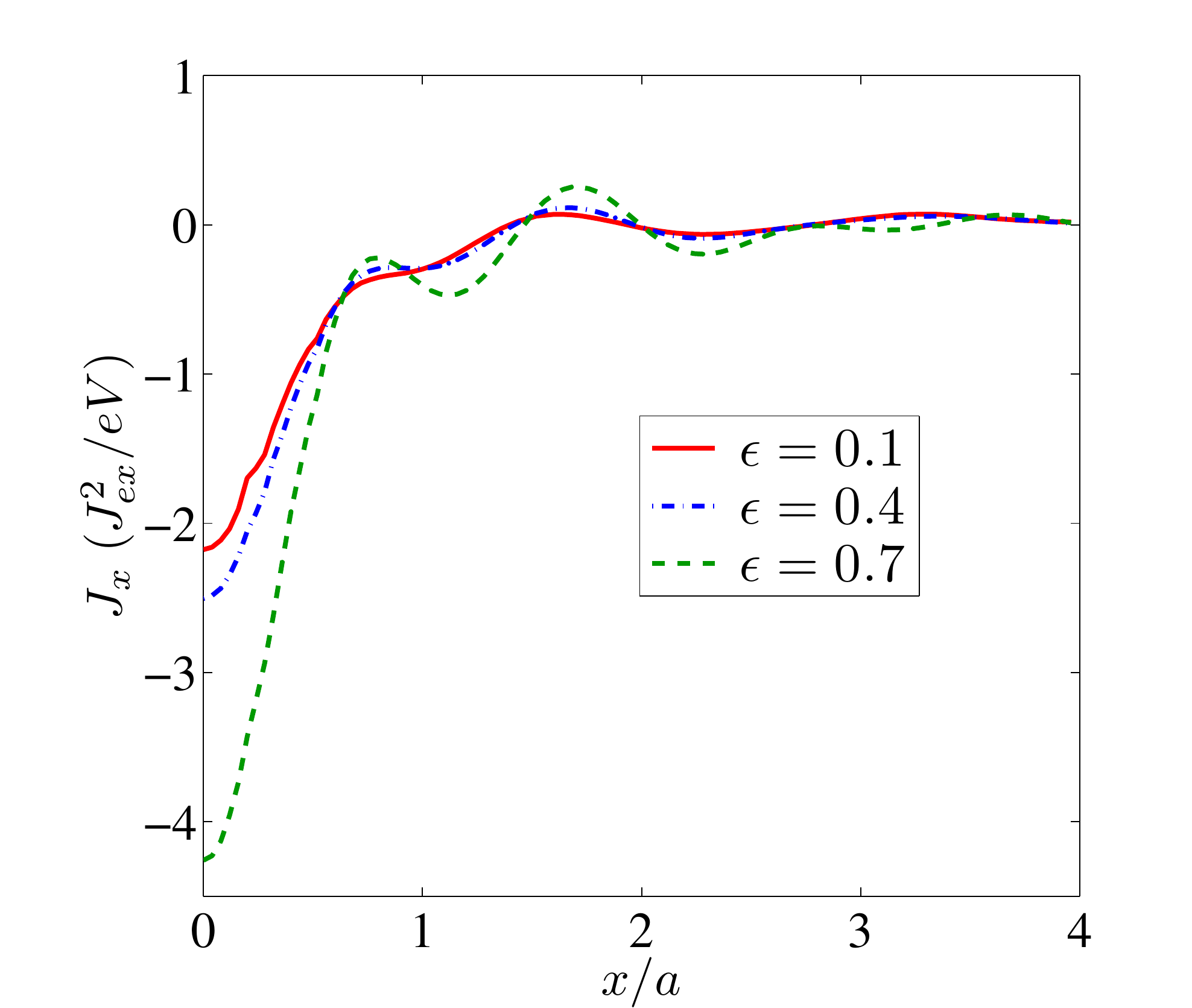}
b)\includegraphics[width=0.4\linewidth,angle=0]{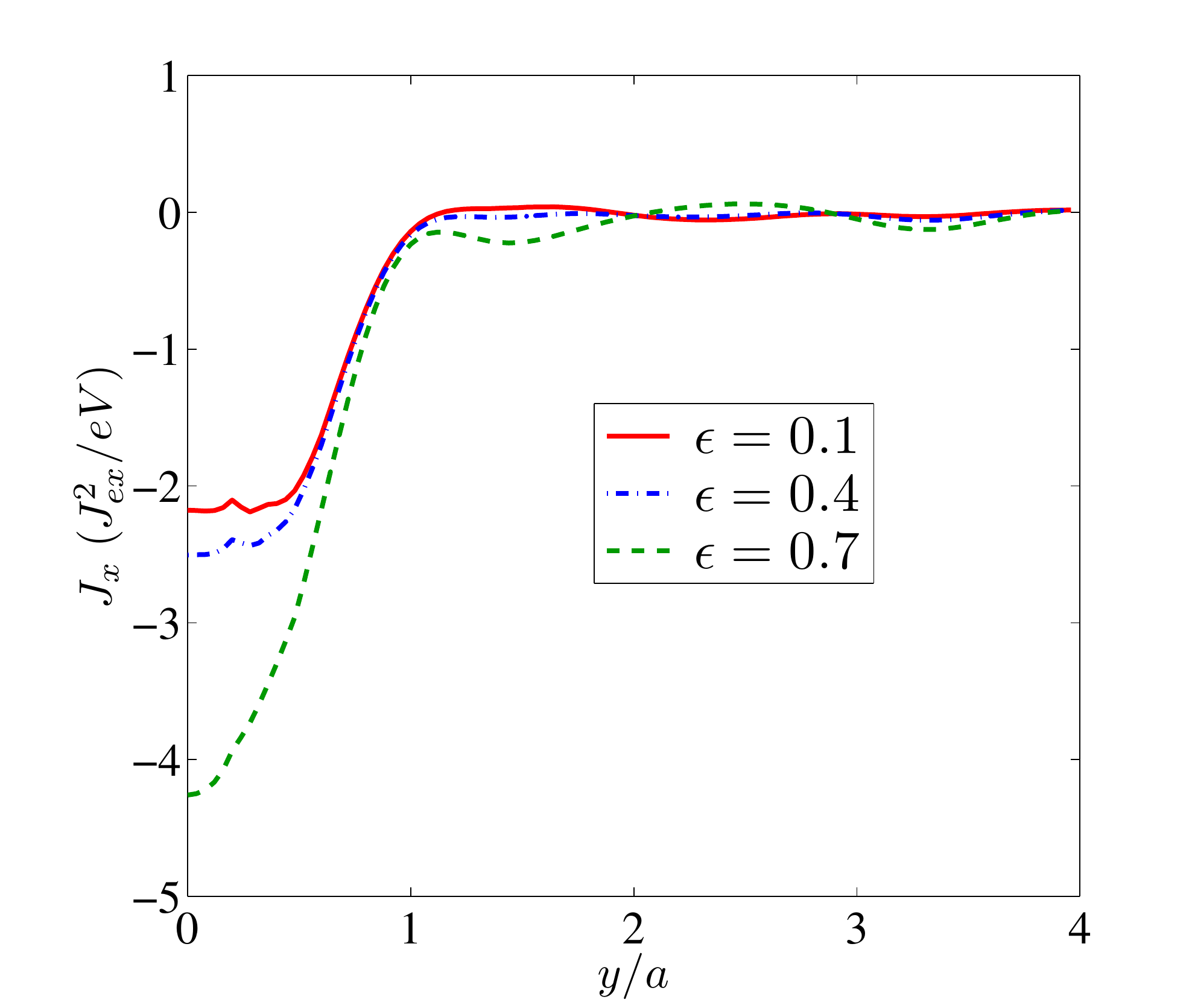}}
\caption{ (color online)
The effect of the ellipticity on the modified XXZ Heisenberg type RKKY interaction in the SDW state, $J_x$,
along $(100)$ (a) - and $(010)$ (b) - crystallographic direction, respectively.}
\label{fig7}
\end{figure*}

The Hamiltonian of the system of localized magnetic moment impurities
in the multi band conduction electron sea is defined by
\bea
{\cal H}={\cal H}_c + {\cal H}_{imp} + {\cal H}_{int},
\eea
where ${\cal H}_c={\cal H}_c^{0}+{\cal H}_c^{\prime}$ is the conduction electron Hamiltonian which is given by:
\bea
{\cal H}_c^{0}&=& \sum_{{\bf  k},\alpha, \sigma}\varepsilon^{\alpha}_{{\bf  k}}
C_{{\bf  k} \alpha \sigma}^\dag C_{{\bf  k}\alpha \sigma}
\nonumber\\
&=&
 \sum_{{\bf  k}, \sigma} \left[ \varepsilon^{h}_{{\bf  k}}
a_{{\bf  k}  \sigma}^\dag a_{{\bf  k} \sigma}
+ \varepsilon^{e_{1}}_{{\bf  k}} b_{1{\bf  k} \sigma}^\dag b_{1{\bf  k} \sigma} +
\varepsilon^{e_{2}}_{{\bf  k}} b_{2 {\bf  k} \sigma}^\dag b_{2{\bf  k} \sigma} \right].
\eea
Here, $C_{{\bf  k} \alpha \sigma}^\dagger$ refers to the creation operators of the conducting electrons. In particular, $a_{{\bf k}\sigma
}^{\dagger}$ ($b_{{\bf k}\gamma\sigma
}^{\dagger}$) creates an electron with spin $\sigma$
in the hole (electron) band. The tight-binding energy dispersion of the electron and hole bands can be parametrized as follows
\bea
 \varepsilon^{h}_{\mathbf{k}} &=&\varepsilon_0 + t_h\left( \cos k_x +\cos k_y \right) -\mu_h\\
\varepsilon^{e_1}_{\mathbf{k}}&= &\varepsilon_0 +
t_e\left( \left[1+\epsilon \right]\cos(k_x+\pi)+\left[ 1-\epsilon \right]\cos(k_y)\right) -\mu_e\nonumber\\
 \varepsilon^{e_2}_{\mathbf{k}}&=& \varepsilon_0 +
t_e\left( \left[1-\epsilon \right]\cos(k_x)+\left[ 1+\epsilon \right]\cos(k_y+\pi) \right)-\mu_e \quad,\nonumber
\eea
where $\epsilon$ accounts for the ellipticity of the electron pockets
and $\varepsilon_0=0.05$eV is the chemical potential for zero doping.
Following our previous analysis\cite{knolle} we use the following hopping matrix elements
$t_h=0.85$ eV, $t_e=-0.68$ eV, $\mu_{h}=1.44$ eV, and $\mu_e=-1.23$ eV which accounts for the Fermi velocities and sizes of the Fermi pockets, see  Ref.~\onlinecite{LDA}.

The interaction part of the conduction electron Hamiltonian contains
density-density interactions between hole and electron bands  which give rise to a SDW order between the hole pocket and one of the electron pocket located around the $(\pi,0)$ point of the BZ\cite{eremin}.
Assuming the experimentally observed ${\bf Q}_{AF}={\bf Q}_1 = (\pi,0)$ SDW ordering wave vector
a standard mean-field decoupling yields the self-consistency condition for the SDW order parameter
${\bf W}\propto \sum_{\bf k} \langle a^\dag_{{\bf  k}\delta} b_{1{\bf  k} \delta^\prime}
\vec{\sigma}_{\delta\delta^\prime} \rangle$. The resulting mean-field Hamiltonian has the form
\bea
{\cal H}_c^{MF}={\cal H}_c^{0}+\sum\limits_{{\bf k} \sigma}
W \sigma
\left[ a^\dag_{{\bf  k}\sigma} b_{1{\bf k}+{\bf Q}_1 \sigma}+ H.c.\right],
\eea
where the spin index $\sigma=\pm 1$  refers to the spin $\uparrow$ and $\downarrow$ respectively.
Now applying the unitary transformations
\bea &&
a_{{\bf  k}\sigma}=v_{{\bf k}}c_{{\bf k}\sigma}-u_{{\bf k}}d_{{\bf k}\sigma}
\nonumber\\&&
b_{1{\bf k}+{\bf Q}_1 \sigma}=\sigma[u_{{\bf k}}c_{{\bf k}\sigma}+v_{{\bf k}}d_{{\bf k}\sigma}]
\eea
the mean-field Hamiltonain for the conducting electrons can be diagonalized and the coefficients of the transformation are given by
\bea && u_{{\bf k}}^2=\frac{1}{2}\left[
1+ \frac{(\varepsilon^{h}_{{\bf  k}}-\varepsilon^{e_1}_{{\bf  k}})}{\sqrt{
(\varepsilon^{h}_{{\bf  k}}-\varepsilon^{e_1}_{{\bf  k}})^2+4W^2 }
}
\right],
%
\nonumber
\\
&&
v_{{\bf k}}^2=\frac{1}{2}\left[
1- \frac{(\varepsilon^{h}_{{\bf  k}}-\varepsilon^{e_1}_{{\bf  k}})}{\sqrt{
(\varepsilon^{h}_{{\bf  k}}-\varepsilon^{e_1}_{{\bf  k}})^2+4W^2 }
}
\right].
\eea
The diagonalized Hamiltonian has the form
\bea
{\cal H}_c^{MF^{(d)}}=
\sum_{{\bf  k}, \sigma} \left[
 E^{1}_{{\bf  k}}
c_{{\bf  k}  \sigma}^\dag c_{{\bf  k} \sigma}
+ E^{2}_{{\bf  k}} d_{{\bf  k} \sigma}^\dag d_{{\bf  k} \sigma}
+E^{ 3}_{{\bf  k}} b_{2 {\bf  k} \sigma}^\dag b_{2{\bf  k} \sigma}
\right],
\nonumber\\
\eea
with quasiparticle energies
\bea
E^{ 1,2}_{{\bf  k}}&=&\frac{1}{2}\left[
(\varepsilon^{h}_{{\bf  k}}+\varepsilon^{e_1}_{{\bf  k}})\pm \sqrt{
(\varepsilon^{h}_{{\bf  k}}-\varepsilon^{e_1}_{{\bf  k}})^2+4W^2 }
\right],
\nonumber\\ E^{ 3}_{{\bf  k}}&= &\varepsilon^{e_{2}}_{{\bf  k}}.
\eea
Note that the electron pocket located around the $(0,\pi)$
point of the BZ remains intact and is not involved
in the SDW formation. In addition, the SDW order explicitly breaks the spin-rotational
symmetry. As a result the RKKY interaction will not be spin-rotationally symmetric and will have a distinct magnetic anisotropy induced in the SDW state.

The interaction part of the local magnetic moments
 with conduction electrons is given by a contact exchange term
\bea
{\cal H}_{int}=-J_{ex}\sum\limits_{{\bf r}; i} s({\bf r}) \cdot {\bf S}_i \;\; \delta({\bf r}-{\bf R}_i)
\eea
where $J_{ex}$ is the exchange coupling constant.
It can be obtained from a more microscopic Anderson type model introduced in
Ref.~\onlinecite{Akbari10}.
Here  ${\bf S}_i$ is the
moment of localized $f$-electrons at site ${\bf R}_i$, and
$s({\bf r})$   is the spin of
conduction electrons.

In the spin density wave state we employ the standard second order perturbation theory with respect to ${\cal H}_{int}$. Its application is straightforward and one finds after some algebra the RKKY interaction which describes the interaction
between two local impurity spins at the positions $i$ and $j$ in the form of
an XXZ type effective exchange Hamiltonian
\bea
{\cal H}_{RKKY}^{  ij}=J^{  ij}_{x}(S_{i}^xS_{j}^x +
S_{i}^yS_{j}^y)
 +J^{  ij}_{z} S_{i}^zS_{j}^z
 \label{eq:RKKYgen}
\eea
The magnetic anisotropy of the form $J_x=J_y\neq J_z$ in this expression appears through
the SDW order which is
  polarized
   along $z$-spin quantization axis.
Specifying  ${\bf R}_i= {\bf 0}$ and ${\bf R}_j= {\bf r}=(x,y)$ these effective
exchange couplings are given by
\bea\nonumber
&&
J^{  ij}_{x}=J^{  ij}_{y}=\sum\limits_{\gamma\gamma^\prime}{\cal J}^{  \gamma\gamma^\prime}_{x}( {\bf r})
=J_{ex}^2\sum\limits_{{\bf k}{\bf k}^\prime\sigma\sigma^\prime\gamma\gamma^\prime}
e^{-i({\bf k}-{\bf k}^\prime)\cdot {\bf r} }
\times\\
&&\hspace{1cm}
\eta_{{\bf k}{\bf k}^\prime\sigma\sigma^\prime}^{i\gamma\gamma^\prime}
{\bf\sigma}_{\sigma\sigma^\prime}^x
\eta_{{\bf k}^\prime{\bf k}\sigma^\prime\sigma}^{j\gamma^\prime\gamma\star}
\left[
\frac{f({\bf k},\gamma)-f({\bf k}^\prime,\gamma^\prime)}
 {E _{{\bf k}}^\gamma-E _{{\bf k}^\prime}^{\gamma^\prime}}
\right],
%
\\\nonumber
&&
J^{  ij}_{z}=\sum\limits_{\gamma\gamma^\prime} {\cal J}^{  \gamma\gamma^\prime}_{z}( {\bf r})
=
J_{ex}^2\sum\limits_{{\bf k}{\bf k}^\prime\sigma\sigma^\prime\gamma\gamma^\prime}
e^{-i({\bf k}-{\bf k}^\prime)\cdot {\bf r} }
\times\\
&&\hspace{1cm}
\eta_{{\bf k}{\bf k}^\prime\sigma\sigma^\prime}^{i\gamma\gamma^\prime}
{\bf\sigma}_{\sigma\sigma^\prime}^z
\eta_{{\bf k}^\prime{\bf k}\sigma^\prime\sigma}^{j\gamma^\prime\gamma\star}
\left[
\frac{f({\bf k},\gamma)-f({\bf k}^\prime,\gamma^\prime)}
 {E _{{\bf k}}^\gamma-E _{{\bf k}^\prime}^{\gamma^\prime}}
\right].
\eea
Here $f({\bf k},\gamma)$ is the Fermi function, and the SDW coherence factors,
 $\eta_{{\bf k}{\bf k}^\prime\sigma\sigma^\prime}^{i\gamma\gamma^\prime}$, are defined as
\bea
&&\eta_{{\bf k}{\bf k}^\prime\sigma\sigma^\prime}^{i11}=
v_{{\bf k}}v_{{\bf k}^\prime}
+\sigma^\prime
e^{i{\bf Q}_1\cdot {\bf R}_i }v_{{\bf k}}u_{{\bf k}^\prime}
+\sigma
e^{-i{\bf Q}_1\cdot {\bf R}_i }u_{{\bf k}}v_{{\bf k}^\prime}
\nonumber\\\nonumber &&\hspace{1.6cm}
+\sigma\sigma^\prime u_{{\bf k}}u_{{\bf k}^\prime}
\\
&&\eta_{{\bf k}{\bf k}^\prime\sigma\sigma^\prime}^{i12}=
-v_{{\bf k}}u_{{\bf k}^\prime}
+\sigma^\prime
e^{i{\bf Q}_1\cdot {\bf R}_i }v_{{\bf k}}v_{{\bf k}^\prime}
-\sigma
e^{-i{\bf Q}_1\cdot {\bf R}_i }u_{{\bf k}}u_{{\bf k}^\prime}
\nonumber\\\nonumber &&\hspace{1.6cm}
+\sigma\sigma^\prime u_{{\bf k}}v_{{\bf k}^\prime}
\\
&&\eta_{{\bf k}{\bf k}^\prime\sigma\sigma^\prime}^{i21}=
-u_{{\bf k}}v_{{\bf k}^\prime}
-\sigma^\prime
e^{i{\bf Q}_1\cdot {\bf R}_i }u_{{\bf k}}u_{{\bf k}^\prime}
+\sigma
e^{-i{\bf Q}_1\cdot {\bf R}_i }v_{{\bf k}}v_{{\bf k}^\prime}
\nonumber\\\nonumber &&\hspace{1.6cm}
+\sigma\sigma^\prime v_{{\bf k}}u_{{\bf k}^\prime}
\\
&&\eta_{{\bf k}{\bf k}^\prime\sigma\sigma^\prime}^{i22}=
u_{{\bf k}}u_{{\bf k}^\prime}
-\sigma^\prime
e^{i{\bf Q}_1\cdot {\bf R}_i }u_{{\bf k}}v_{{\bf k}^\prime}
-\sigma
e^{-i{\bf Q}_1\cdot {\bf R}_i }v_{{\bf k}}u_{{\bf k}^\prime}
\nonumber\\\nonumber &&\hspace{1.6cm}
+\sigma\sigma^\prime v_{{\bf k}}v_{{\bf k}^\prime}
\\\nonumber
&&\eta_{{\bf k}{\bf k}^\prime\sigma\sigma^\prime}^{i13}=
v_{{\bf k}}e^{i{\bf Q}_2\cdot {\bf R}_i }
+\sigma
u_{{\bf k}}
e^{-i({\bf Q}_1-{\bf Q}_2)\cdot {\bf R}_i }
\\\nonumber
&&\eta_{{\bf k}{\bf k}^\prime\sigma\sigma^\prime}^{i31}=
v_{{\bf k}^\prime}e^{-i{\bf Q}_2\cdot {\bf R}_i }
+\sigma^\prime
u_{{\bf k}^\prime}
e^{i({\bf Q}_1-{\bf Q}_2)\cdot {\bf R}_i }
\\
&&\eta_{{\bf k}{\bf k}^\prime\sigma\sigma^\prime}^{i23}=
-u_{{\bf k}}e^{i{\bf Q}_2\cdot {\bf R}_i }
+\sigma
v_{{\bf k}}
e^{-i({\bf Q}_1-{\bf Q}_2)\cdot {\bf R}_i }
\nonumber\\
&&\eta_{{\bf k}{\bf k}^\prime\sigma\sigma^\prime}^{i32}=
-u_{{\bf k}^\prime}e^{-i{\bf Q}_2\cdot {\bf R}_i }
+\sigma^\prime
v_{{\bf k}^\prime}
e^{i({\bf Q}_1-{\bf Q}_2)\cdot {\bf R}_i }
\nonumber\\
&&\eta_{{\bf k}{\bf k}^\prime\sigma\sigma^\prime}^{i33}=
1
\eea
Setting $W=0$ it is easy to verify that  in the paramagnetic or normal state regime the
 RKKY interaction simplifies to the usual expression in two-dimensional  metals with nesting properties\cite{Aristov97a}
\bea
{\cal H}_{RKKY}^{  ij}=J_n^{  ij}{\bf S}_{i}\cdot {\bf S}_{j}
\eea
where the interaction is now isotropic in the spin space ($J_x =J_y=J_z \equiv J_n$).
The effective exchange couplings are then given by
\bea
J_n^{  ij}=\sum\limits_{\gamma\gamma^\prime} {\cal J}^{  \gamma\gamma^\prime}_{n}({\bf r})=J_{ex}^2
Re\left(\sum\limits_{\gamma\gamma^\prime}
e^{i({\bf Q}_{\gamma}-{\bf Q}_{\gamma^\prime})\cdot {\bf r}}  \chi^{\gamma\gamma^\prime} ({\bf r})\right),
\nonumber\\
\eea
here ${\bf Q}_{e_1}={\bf Q}_{1}$, ${\bf Q}_{e_2}={\bf Q}_{2}$, ${\bf Q}_{h}={0}$, and
$\chi^{\gamma\gamma^\prime} ({\bf r})$ is magnetic spin susceptibility
of conduction electrons (Lindhard response function) which is given by
\bea
\chi^{\gamma\gamma^\prime} ({\bf r})=
\sum\limits_{{\bf k}{\bf k}^\prime}
e^{i({\bf k}-{\bf k}^\prime)\cdot {\bf r}}
\left[
\frac{f({\bf k},\gamma)-f({\bf k}^\prime,\gamma^\prime)}
 {\varepsilon _{{\bf k}}^{\gamma}-\varepsilon _{{\bf k}^\prime}^{\gamma^\prime}}
\right].
\eea
%


\begin{figure*}
\centerline{
a)\includegraphics[width=0.4\linewidth,angle=0]{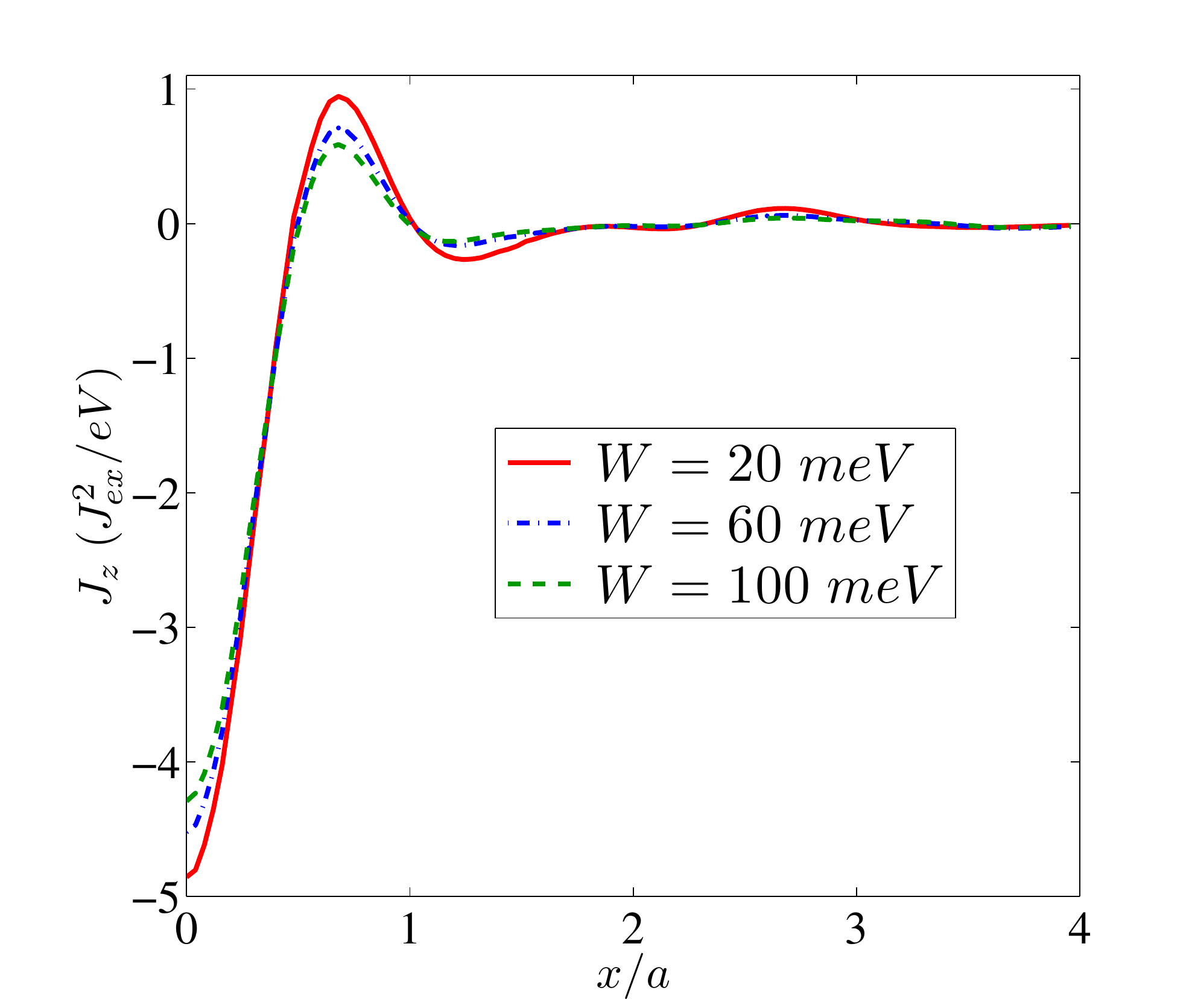}
b)\includegraphics[width=0.4\linewidth,angle=0]{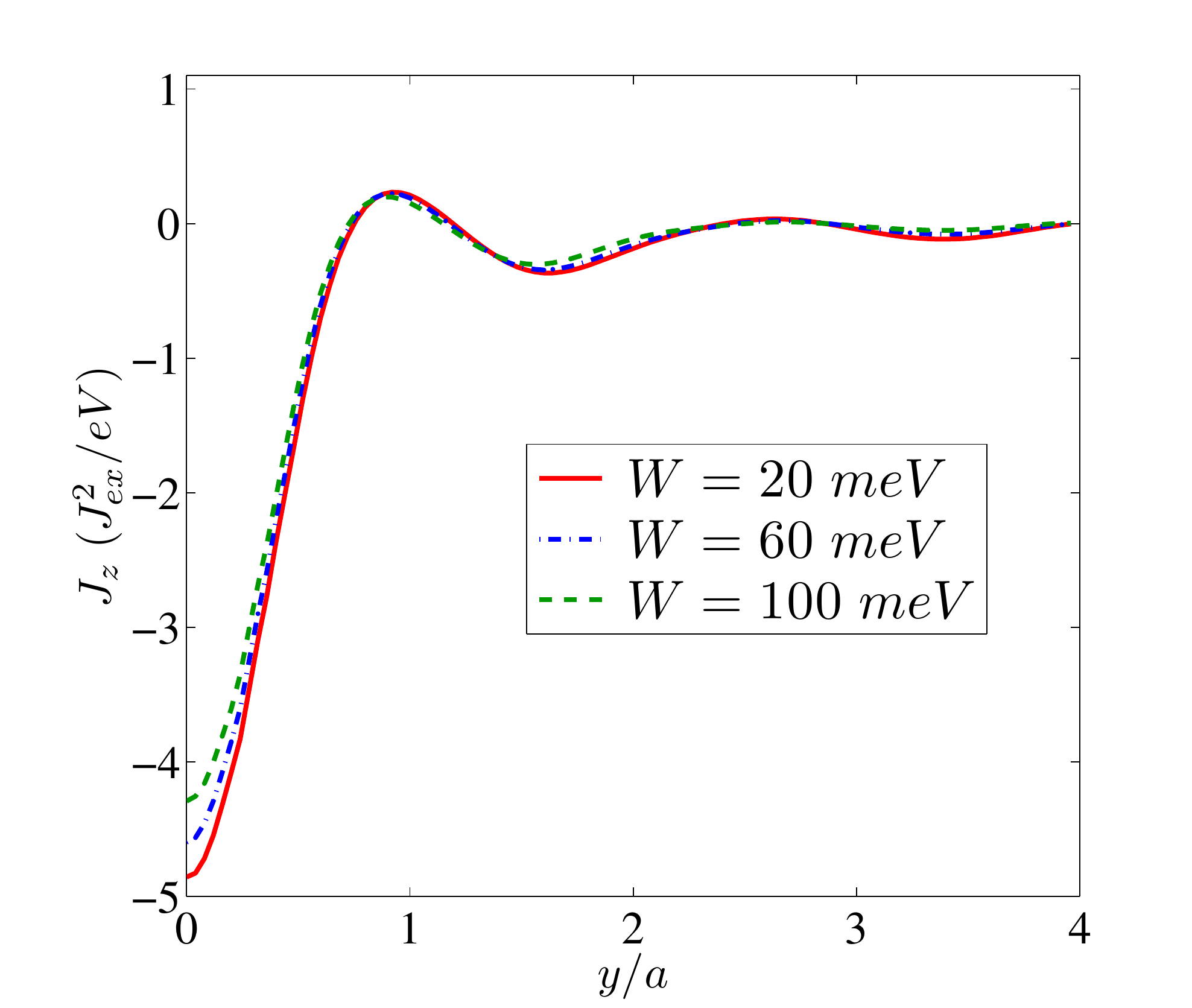}}
\caption{ (color online)
Influence of the SDW gap magnitude on the RKKY interaction parameter, $J_z$,
for $(100)$-direction (a); and for $(010)$-direction (b).}
\label{fig8}
\end{figure*}

 \section{Numerical results and discussion}

Based on the equations above  we present in the following
 the results of the numerical evaluation of the RKKY interaction in the SDW and the normal state phases.
We first present an overall behavior of the RKKY interaction. In particular, Fig.(\ref{fig2}) shows the contour mesh of the RKKY interaction for $5\%$ electron doping and ellipticity parameter $\epsilon=0.5$ as a function of interatomic distances for both SDW and normal state regimes.
In addition to the breaking of the spin rotational symmetry, there is another important difference in the behavior of the
RKKY interaction between the normal and the SDW state of iron-based superconductors. This concerns the absence of tetragonal symmetry in the SDW
state. As clearly seen by comparing Fig.(\ref{fig2})(a) and Fig.(\ref{fig2})(b)-(c) the $C_4$ symmetry present in the normal state is broken down to $C_2$ symmetry. Apart from these differences the RKKY interaction show also some similarities. In particular,
close to the  impurity position the effective interaction is ferromagnetic(FM)
in both cases and becomes antiferromagnetic(AF) at a distance comparable to
the lattice constant. In the asymptotic regime ($x/a, y/a \gg 1$) the oscillatory behavior from FM to AF and vice versa sets in.
As expected the amplitude of oscillations decreases with increasing distance between the local moment impurities.

For clarification of the role played by the interband and intraband scattering
we display in Fig.~\ref{fig3} the contribution of each term separately in the normal state.
These figures show that the inter band contribution to the RKKY interaction is much larger at shorter distances than the intraband one. This arises again due to pronounced nesting features of the electron and hole bands. In addition, observe that a finite ellipticity introduces some asymmetry along $x$ and $y$ direction  for each component of the RKKY interaction where electron pockets are involved. This asymmetry, however, averages out in the full RKKY interaction which possesses again the tetragonal symmetry.

This is not any longer the case in the SDW state as shown in Fig.\ref{fig4} for $J_x$ component. Here, the intra- and inter-band contributions to the RKKY  are strongly anisotropic. This originates from the fact that the SDW has ${\bf Q}_1 = (\pi,0)$ ordering wave vector and as a result of its ordering the tetragonal symmetry is broken. It is interesting to notice that such effect was found recently in EuFe$_2$As$_2$ where the magnetic anisotropy of the Eu magnetic moments was changing across the SDW transition temperature\cite{zapf}. This change in the magnetic anisotropy is in direct agreement with our results. Note also that the sum of
all contributions is anisotropic along $a$ and $b$ crystallographic directions in both $J_x$ and $J_z$ components of the interactions. In particular, as clearly shown in Fig.\ref{fig2}(b)-(c) the ferromagnetic interaction is extended in both  $J_x$ and $J_z$ along the $y$ direction  which is perpendicular to the AF ordering of the Fe-plane.

To see the changes on the quantitative level we show in Fig.\ref{fig5} the dependence of the RKKY interaction along $x$ direction in both normal and SDW states. We observed  that the amplitude of the oscillations increases in the SDW state but overall the dependence remains the same. Namely, it is ferromagnetic for short distances and then oscillates between positive (AF) and negative (FM) values. In addition, $J_z$ and $J_x$ RKKY interactions have slightly different period which results in the fact that they may have opposite signs for a given inter-impurity distance.
This difference is associated with the structure of the SDW matrix elements which appear in the $J^z$ and $J^x$ components of the RKKY interaction differently (see Eqs.(11)-(12)). An additional effect of the SDW is shown in Fig.\ref{fig6} where we plot
the behavior of $J_z$ and $J_x$ along $a$ and $b$ crystallographic directions. As clearly seen from this figure, the period of the oscillation is not only different for $J_x$ and $J_z$ components but also for each of them in the $x$ and $y$ crystallographic directions. In particular, the antiferromagnetic  XXZ regime with negative anisotropy at $x \simeq 2a$ is observed along $(0,1)$ direction.
On the other hand for (1,0)-direction the XXZ ferromagnetic Heisenberg model is
dominant and the effective interaction changes to AF behavior  only around  $y \simeq 2.5 a$.
  Thus through SDW spin-space and real space anisotropies are correlated.
Furthermore along (1,0)-direction the magnetic anisotropy changes sign for the first time at a distance of about $y \simeq 0.75 a$ and it
prevails for a longer period as compared to (0,1) direction. The nature of this difference is both SDW order and the structure of the remaining small pockets that occur due to folding of hole and one electron pocket located at $(\pi,0)$ point of the BZ. Due to larger hopping of this electron band
along $x$-direction the $k_F$ values of the folded bands in the SDW state are unequal  along $x$ and $y$-direction which is then reflected in the periodicity of the RKKY interaction.

  Furthermore we show the effect of the ellipticity
   in Fig.\ref{fig7} where one could clearly observe the increasing period of the oscillations for larger values of $\epsilon$. The same effect is observed for increasing electron doping. This is natural as both ellipticity and doping for a given value of $W$ make the remnant pockets and the corresponding values of $k_F$ along $x$ and $y$ direction larger. This effect is almost absent in the normal state (not shown) which points out that in the SDW state the interaction between magnetic impurities will be strongly modified.

The natural question arises whether the effects of the SDW state on the RKKY interaction become more pronounced for increasing size of the SDW gap and corresponding increase of the magnetic moment. In Fig.\ref{fig8} we show the evolution of the oscillatory behavior of the RKKY $J_x$ interaction for different value of $W$. Note that the amplitude of the oscillations weakens upon increase of the SDW gap. This is due to the shrinkage of the
remnant electron and hole FS pockets which arise due to folding of the BZ in the SDW. The larger becomes the SDW gap the smaller will be the remnant pocket size. As a matter of fact for some critical value of $W$ the pockets involved in the SDW completely disappear from the Fermi surface. Therefore the only contribution to the RKKY will arise in this case due to electron pocket located at $(0,\pi)$, not involved in the SDW formation. This explains why $J_z$ shows weaker oscillations along $x$-direction, while along $y$-direction the oscillations are almost the same. This is because the oscillatory behavior of $J_z$ along $(01)$ direction is determined by the electron pocket which remains intact in the SDW state while along $(1,0)$ direction the SDW order gaps completely the FS and only slight oscillations are still visible.

 \section{Summary and Conclusion}

In conclusion we analyze the changes of the RKKY interaction in the SDW state of iron-based superconductors.
The generalized RKKY interaction in these compounds  is of an effective XXZ Heisenberg-type
where the $O(3)$ symmetry is broken but $U(1)$ symmetry of the interaction for rotation around an
axis parallel to the SDW polarization vector $ W {\hat z}$ is still preserved.
We show that for small  distances between the local moments, $r<a$, the interaction between local spins is ferromagnetic but
for larger $r$ it oscillates between AF and FM regimes with different periods and amplitudes which depend strongly
on ellipticity or doping  of electron pockets. In addition, the period of the oscillation strongly depends on the magnitude of the SDW gap
and the structure of the Fermi surface in the folded BZ. 

Our main observation is that  the RKKY interaction between magnetic impurities in SDW state become anisotropic
below T$_{SDW}$. As a result, the magnetization of the rare-earth magnetic
moments, already anisotropic by itself due to crystalline electric field effects, will experience additional temperature dependent
anisotropy induced by the conduction  electrons  below T$_{SDW}$. Quite generally the effect of SDW ordering of Fe spins
on the rare-earth subsystem was found in several studies\cite{maeter,nandi}. However, the effect of induced anisotropy below T$_{SDW}$
on the rare-earth magnetization was observed only  recently in EuFe$_2$(As$_{1-x}$P$_x$)$_2$ system by measuring
magnetic anisotropy of the Eu$^{2+}$ ions above and below T$_{SDW}$. In particular, it was found that upon
decreasing temperature the ratio the magnetization anisotropy of Eu spins, $M_{ab}/M_c$, becomes temperature dependent below T$_{SDW}$ reflecting the influence of the SDW order\cite{zapf}. This is in direct agreement with our results.
Our further observation that the magnetic anisotropy is then also reflected in the spatial anisotropy was not yet observed as it requires the use of untwinned
crystals or the use of the local probes such as nuclear magnetic resonance (NMR). It would be interesting to check this effect experimentally.
\\

 \section*{Acknowledgments}
We would like to acknowledge S. Zapf for useful discussion and sharing with us the experimental results prior to publication. IE is thankful to Kazan Federal University 
(Grant RNP-31) for the partial support.




\begin{thebibliography}{99}
\bibitem{Ruderman54}
M.A. Ruderman and C. Kittel, Phys. Rev. {\bf 96}, 99 (1954).

\bibitem{Kasuya56}T. Kasuya, Prog. Theor. Phys. {\bf 16}, 45 (1956).
 %
 \bibitem{Yosida57}
 K. Yosida, Phys. Rev. {\bf 106}, 893 (1957).
%
 \bibitem{Fisher75}B. Fisher and M. Klein, Phys. Rev. B {\bf 11}, 2025 (1975).
 %
 \bibitem{Monod87}M. T. Beal-Monod,  Phys. Rev. B {\bf 36}, 8835 (1987).
%
 \bibitem{Yafet87}Y. Yafet, Phys. Rev. B {\bf 36}, 3948 (1987).
 %
%
%
 \bibitem{Aristov97a} D. N. Aristov, and S. V. Maleyev, Phys. Rev. B {\bf  56}, 8841 (1997).
 %
 %
\bibitem{Kamihara08} Y. Kamihara, T. Watanabe, M.  Hirano and  H. Hosono, J. Am. Chem. Soc. {\bf 130}, 3296 (2008 ).

\bibitem{LDA} S. Lebegue
Phys. Rev. B {\bf 75}, 035110 (2007); D.J. Singh and M.-H. Du,
 Phys. Rev. Lett. \textbf{100}, 237003 (2008);
 L. Boeri, O.V. Dolgov, and A.A. Golubov, Phys.
 Rev. Lett. \textbf{101}, 026403 (2008);
 I.I. Mazin, D.J. Singh, M.D. Johannes, and M.H. Du,
 Phys. Rev. Lett. \textbf{101}, 057003 (2008).
%

\bibitem{ARPES} C. Liu, G.D. Samolyuk, Y. Lee, N. Ni, T. Kondo, A.F. Santander-Syro, S.L. Bud'ko, J.L. McChesney, E. Rotenberg, T. Valla, A. V. Fedorov, P.C. Canfield, B.N. Harmon, A. Kaminski,
 Phys. Rev. Lett. {\bf 101}, 177005 (2008);  D.V. Evtushinsky, D.S. Inosov, V.B. Zabolotnyy, A. Koitzsch, M. Knupfer, B. B\"uchner, M.S. Viazovska, G.L. Sun, V. Hinkov, A.V. Boris, C.T. Lin, B. Keimer, A. Varykhalov, A.A. Kordyuk, and S.V. Borisenko, Phys. Rev. B {\bf 79}, 054517 (2009);  D. Hsieh, Y. Xia, L. Wray, D. Qian, K. Gomes, A. Yazdani, G.F. Chen, J.L. Luo, N.L. Wang, and M.Z. Hasan,
 arXiv:0812.2289 (unpublished);  H. Ding, K. Nakayama, P. Richard, S. Souma, T. Sato, T. Takahashi, M. Neupane, Y.-M. Xu, Z.-H. Pan, A.V. Federov, Z. Wang, X. Dai, Z. Fang, G.F. Chen, J.L. Luo, N.L. Wang,
J. Phys.: Condens. Matter 23, 135701 (2011).

\bibitem{coldea}
A.I. Coldea, J.D. Fletcher, A. Carrington, J.G. Analytis, A.F. Bangura, J.-H. Chu, A.S. Erickson, I.R. Fisher, N.E.
Hussey, and R.D. McDonald, Phys. Rev. Lett. {\bf 101}, 216402 (2008); J. G. Analytis, C. M. Andrew, A. I. Coldea, A. McCollam, J.-H. Chu, R. D. McDonald, I. R. Fisher, and A. Carrington
Phys. Rev. Lett. 103, 076401 (2009).

\bibitem{sebastian} S.E. Sebastian, J. Gillett, N. Harrison, P.H.C. Lau,
D.J. Singh, C.H. Mielke, and G.G. Lonzarich, J. Phys. Condens. Matter {\bf 20} 422203 (2008).

\bibitem{Tesanovic} V. Cvetkovic and Z. Tesanovic, EPL {\bf 85}, 37002 (2009).
See also  V. Stanev, J. Kang, and Z. Tesanovic, Phys. Rev. B 78, 184509 (2008).

\bibitem{Chubukov2008} A.V. Chubukov, D.V. Efremov, and I. Eremin, Phys. Rev. B \textbf{78}, 134512 (2008);
A.V. Chubukov, Physica C {\bf 469}, 640 (2009).


\bibitem{d_h_lee} Fa Wang, Hui Zhai, Ying Ran, Ashvin Vishwanath, and Dung-Hai Lee, Phys. Rev. Lett. {\bf 102}, 047005 (2009).
\bibitem{Korshunov2008} M.M. Korshunov and I. Eremin, Phys. Rev. B
    \textbf{78}, 140509(R) (2008); Europhys. Lett. \textbf{83}, 67003 (2008).
\bibitem{timm} P.M.R. Brydon and C. Timm, Phys. Rev. B {\bf 79}, 180504(R) (2009).
\bibitem{honerkamp} C. Platt, C. Honerkamp, and W. Hanke, New J. Phys. {\bf 11}, 055058 (2009).

\bibitem{eremin} I. Eremin and A.V. Chubukov, Phys. Rev. B {\bf 81}, 024511 (2010).

\bibitem{maeter} H. Maeter, H. Luetkens, Yu.G. Pashkevich, A. Kwadrin, R. Khasanov, A. Amato, A.A. Gusev,
K.V. Lamonova, D.A. Chervinskii, R. Klingeler, C. Hess, G. Behr, B. B\"uchner, and H.-H. Klauss, Phys. Rev. B {\bf 80} 094524 (2009); A. Jesche, C. Krellner, M. de Souza, M. Lang, and C. Geibel, New Jour. Phys. {\bf 11}, 103050 (2009).

\bibitem{pourovskii} L. Pourovskii, V. Vildosola, S. Biermann, and A. Georges, Europhys. Lett. {\bf 84}, 37006 (2008).

\bibitem{si} Q. Si, E. Abrahams, J. Dai, and J.-X. Zhu, New J. Phys. {\bf 11}, 045001 (2009); A.H. Nevidomskyy, and P. Coleman, Phys. Rev. Lett. {\bf 103}, 147205 (2009).

\bibitem{knolle} J. Knolle, I. Eremin, A. V. Chubukov, and R. Moessner, Phys. Rev. B {\bf 81}, 140506(R) (2010).
 %
%
\bibitem{Akbari10} A. Akbari, I. Eremin and P. Thalmeier, Phys. Rev. B {\bf 81}, 014524 (2010).
%
\bibitem{zapf} S. Zapf, D. Wu, L. Bogani, H.S. Jeevan, Ph. Gegenwart, and M. Dressel, arXiv:1103.2446 (unpublished); S. Zapf, private communication.

\bibitem{nandi} S. Nandi, Y. Su, Y. Xiao, S. Price, X. F. Wang, X. H. Chen, J. Herrero-Martin, C. Mazzoli, H. C. Walker, L. Paolasini, S. Francoual, D. K. Shukla, J. Strempfer, T. Chatterji, C.M.N. Kumar, R. Mittal, H. M. R\"onnow, Ch. R\"uegg, D. F. McMorrow, and Th. Br\"uckel,
    Phys. Rev. B {\bf 84}, 054419 (2011).

%



\end{thebibliography}
\end{document}